\documentclass[10pt]{article}
\usepackage{natbib}
\usepackage{amsmath,amssymb,amsthm}
\usepackage{float}
\usepackage{footnote}
\usepackage{subcaption}
\usepackage{bbm}
\usepackage{algorithm}
\usepackage[noend]{algpseudocode}
\usepackage{mathtools}% http://ctan.org/pkg/mathtools

\def\NoNumber#1{{\def\alglinenumber##1{}\State #1}\addtocounter{ALG@line}{-1}}

\newcounter{algsubstate}
\renewcommand{\thealgsubstate}{\roman{algsubstate}}
\newenvironment{algsubstates}
  {\setcounter{algsubstate}{0}%
   \renewcommand{\State}{%
     \stepcounter{algsubstate}%
     \Statex {\footnotesize\thealgsubstate:}\space}}
  {}

\usepackage{setspace}
\usepackage{authblk}

\usepackage{url}
\usepackage{todonotes}

\usepackage{tikz}

\newtheorem{assumption}{Assumption}

\theoremstyle{definition}

\newcommand\independent{\protect\mathpalette{\protect\independenT}{\perp}}
\def\independenT#1#2{\mathrel{\rlap{$#1#2$}\mkern2mu{#1#2}}}

\providecommand{\keywords}[1]
{
  \small	
  \textbf{\textit{Keywords---}} #1
}

\title{Flexible machine learning estimation of conditional average treatment effects: a blessing and a curse}

\author[1, 2]{R.A.J. Post}
\author[1]{I.L. van den Heuvel}
\author[1]{M. Petkovic}
\author[1, 3]{E.R. van den Heuvel}
\affil[1]{Department of Mathematics and Computer Science, Eindhoven University of Technology, The Netherlands}       
\affil[2]{Institute for Complex Molecular Systems, Eindhoven University of Technology, The Netherlands}    
\affil[3]{Department of Preventive Medicine and Epidemiology, School of Medicine, Boston University, USA}                     
\date{}
\begin{document}
%TC:ignore
		\maketitle

%\newpage
		\abstract{
Causal inference from observational data requires untestable identification assumptions. If these assumptions apply, machine learning (ML) methods can be used to study complex forms of causal effect heterogeneity. Recently, several ML methods were developed to estimate the conditional average treatment effect (CATE). If the features at hand cannot explain all heterogeneity, the individual treatment effects (ITEs) can seriously deviate from the CATE. In this work, we demonstrate how the distributions of the ITE and the CATE can differ when a causal random forest (CRF) is applied. We extend the CRF to estimate the difference in conditional variance between treated and controls. If the ITE distribution equals the CATE distribution, this estimated difference in variance should be small. 
If they differ, an additional causal assumption is necessary to quantify the heterogeneity not captured by the CATE distribution. The conditional variance of the ITE can be identified when the individual effect is independent of the outcome under no treatment given the measured features. Then, in the cases where the ITE and CATE distributions differ, the extended CRF can appropriately estimate the variance of the ITE distribution while the CRF fails to do so.
	 }  \hspace{10pt}    
	 
%TC:endignore

\section{Introduction}
The increasing availability of (big) observational data has tremendously boosted the field of machine learning (ML) \citep{Mooney2018}. ML provides us with flexible, non-parametric methods to study the observed outcome $Y$, given features $\boldsymbol{X}$, that may involve a treatment (or exposure) $A$, by statistical inference on the (conditional) distributions of $Y \mid \boldsymbol{X}, A{{=}}a$. Therefore, ML methods are excellent at predicting future observations that arise from the same (factual) distribution \citep{Dickerman2020}. However, it is essential to realize that these models cannot be automatically used to answer `what if' questions for the treatment $A$, i.e.~for counterfactual prediction, as associations found in the data are not necessarily causal \citep{Hernan2019b, Dickerman2020, Prosperi2020, VanGeloven2020, Mooney2022, Cui2022, Dickerman2022}. Statistical inference of associations is thus only one step in causal inference and, as such, in counterfactual prediction \citep{Balzer2021}. 

The critical step for causal inference is linking the distribution of outcomes in a universe where everyone was treated with $a$, i.e.~potential outcomes \citep{Hernan2019},  $Y^{a} \mid \boldsymbol{X}$ to the distribution of the observed data.  When working with observational data, we have to make assumptions for the identification of causal quantities of interest that cannot be verified with the data, so ML is insufficient. Instead, we have to rely on the knowledge of experts \citep{Hernan2019}. Suppose these identification assumptions can be made so that the distributions of potential and observed outcomes can be linked. In that case, causal estimands (targets) of interest can be connected to estimands of the data-generating distribution and estimated with statistical inference. Next to the validity of the identification assumptions,  accurate statistical inference is thus necessary for causal inference. When the assumptions are applicable, but the statistical inference is off, e.g., when using misspecified models for the observed outcomes, the causal inference will also be invalid. The flexibility offered by ML methods can thus improve statistical inference \citep{Mooney2022, Blakely2019}.   More precisely, ML methods can be exploited to learn nuisance parameters of the data generating distribution, such as conditional means and propensity scores, which in turn can be used to estimate the causal estimand, as is, for example, done in targeted maximum likelihood estimation (TMLE) \citep{Laan2011, Schuler2017}.

The increasing availability of diverse data makes studying effect heterogeneity among individuals more feasible. The field of precision medicine aims to understand this heterogeneity to improve individual treatment decisions \citep{Kosorok2019}. The average treatment effect (ATE), $\mathbb{E}[Y^{1}-Y^{0}]$, might seriously differ from the individual treatment effect (ITE), i.e.~ the actual change in outcome caused by the exposure for a particular individual $i$ ($Y_{i}^{1}-Y_{i}^{0}$) \citep{Kravitz2004}. However, it is well known that an ITE is not identifiable because of the fundamental problem of causal inference \citep{Holland1986}, i.e.~ it is impossible to observe the different potential outcomes for one individual jointly.  
On the other hand, marginalized effects like the ATE and the conditional average treatment effect (CATE) become identifiable in the absence of unmeasured confounding. In randomized experiments, this unconfoundedness assumption holds by design. Treatment effect heterogeneity studies thus focus on the estimation of the individualized CATEs, $\mathbb{E}[Y^{1}-Y^{0} \mid \boldsymbol{X}]$, given measured features $\boldsymbol{X}$, but aggregated over remaining unmeasured features, as a proxy for the ITEs \citep{Robertson2020}. The functional form of effect modification by different levels of the measured features might be very complex, so ML methods are promising tools for estimating CATEs \citep{Bica2021}. 

In recent years several meta-learning strategies for CATE estimation have been proposed. These strategies decompose the CATE estimation into regression problems that can be solved with any suitable ML method (see \citet{Caron2022} for a detailed review).  T-learners fit separate models for treated and controls and estimate CATEs as the plug-in difference of the conditional mean estimates (see, e.g.~\citet{Athey2016} and \citet{Powers2018}). The performance of T-learners will depend on the levels of sparsity and smoothness of conditional means for treated and controls, as well as the choice of the base learner. T-learners generally fail for subgroups where the treated and control samples differ in size, as illustrated by \citet{Kunzel2019}. X-learners have been proposed to deal with the difference in sample size by first using a T-learner to predict individual treatment effects that are subsequently used to derive the CATEs for treated and controls separately from which a weighted average is derived \citep{Kunzel2019}. S-learners include treatment assignment as another covariate next to other features, and the CATE is estimated as the difference of the estimated conditional means for treated and controls (see e.g.~\citet{Hill2011}, \citet{Foster2011}, \citet{Green2012} and \citet{Imai2013}). Estimation with S-learners might suffer from serious finite-sample bias because they do not involve the CATE directly but focus on the conditional means, a problem also known for ATE estimation \citep{Chernozhukov2018}. To remedy this issue, \citet{Hahn2020} introduced the Bayesian causal forest model that extends the work of \citet{Hill2011} by including the CATE as an explicit parameter in the model with its own prior. The R-learner directly identifies the CATE by regressing transformed outcomes on transformed treatment assignment using estimates of nuisance parameters in a first step (as we will elaborate on in Section \ref{CH2CH2:sec2}) \citep{Nie2020}. The R-learner is also called `double machine learning' and may give unbiased estimates of the average causal effect for finite samples. At the same time, a one-step approach (S-learner) would still be biased \citep{Chernozhukov2018}.  Similarly, the DR-learner deals with augmented inverse probability weighted transformation \citep{Robins1994} of observations after constructing estimates of the propensity score and conditional means in a first step \citep{Kennedy2020optimal, Fan2022}.  The cost of making weaker modelling assumptions using the flexible ML methods is slower convergence rates for the estimators, known as the curse of dimensionality \citep{Naimi2022}. Therefore, much of the ongoing research is focused on comparing the different methods for CATE estimation to derive whether and when they are optimal (see e.g.~\citet{Wendling2018}, \citet{Knaus2020}, \citet{Kennedy2020optimal}, \citet{Curth2021} and \citet{kennedy2023minimax}). 

The aim of this work is different, as we want to emphasize the difference between the CATE and the ITE. The CATE is much more personalized than the ATE and, thus, an important step towards precision medicine. However, it concerns us that the CATE is sometimes perceived as equivalent to the ITE (see, e.g.~\citet{Lu2018}). Whether the CATE can appropriately approximate the ITE depends on the remaining variability of causal effects given the considered modifiers, e.g.~a CATE $\geq 0$ given $\boldsymbol{X}{=}\boldsymbol{x}$ \citep{Talisa2021} does not imply that all ITEs $\geq 0$ for those individuals \citep{Hand1992}. In this work, we investigate whether we can use a causal random forest (CRF) \citep{Athey2019} to estimate the variance of the marginal ITE distribution. More specifically, we investigate the performance of the CRF to estimate $\text{var}(Y^1-Y^0\mid \boldsymbol{X}{=}\boldsymbol{x})$ and $\text{var}(Y^1-Y^0)$ %and $\mathbb{P}(Y^1-Y^0>0)$ 
that could be essential characteristics next to the ATE and CATE. To do so, we have simulated data from a causal system based on estimates from a real case study and fit the CRF to estimate individual CATEs and compared the distributions of the (random) conditional expectation, $\mathbb{E}[Y^1-Y^0  \mid \boldsymbol{X}]$, and the ITE, $Y^1-Y^0$.     

To open up the field of ITE distribution estimation, we derive what identification assumption should hold, additionally to those necessary for marginal causal inference, to identify other characteristics of the conditional ITE distribution. We show that under a conditional independent effect deviation assumption (ITE independent of $Y^{0}$ given the CATE), the (conditional) variance of the ITE becomes identifiable. This new identifiability assumption can also not be verified with data, but contrary to the unconfoundedness assumption, the assumption might be violated in a randomized experiment. To give an idea of how an assumption on the joint distribution of potential outcomes can evolve the field of treatment effect heterogeneity, we extend the CRF algorithm to estimate the variance of the ITE given the measured features. Suppose one would additionally be willing to assume a Gaussian distribution. In that case, the algorithm outputs a conditional ITE distribution (centred at the CATE) that will be degenerate in the absence of remaining effect heterogeneity. 

Section \ref{CH2CH2:sec2} introduces our notation and presents the identification assumptions necessary for CATE estimation. Furthermore, we describe the methodology behind the CRF algorithm and our reality-based data simulation. Section \ref{CH2CH2:sec3} presents the results of fitting the CRF to datasets simulated under different settings. In Section \ref{CH2CH2:sec4}, we introduce the new causal assumption such that the conditional variance of the ITE becomes identifiable and extend the CRF to estimate this. Furthermore, we present the results of analyzing the simulated datasets with this new algorithm. Finally, we present some concluding remarks and ideas for future research in Section \ref{CH2CH2:sec5}. 

\section{Notation and Methods}\label{CH2CH2:sec2}
Probability distributions of factual and counterfactual outcomes are defined in the potential outcome framework \citep{Neyman1990, Rubin1974}. Let $Y_{i}$ and $A_{i}$ represent the (factual) stochastic outcome and the random treatment assignment level of individual $i$. Let $Y_{i}^{a}$ equal the potential outcome under an intervention on the treatment to level $a$ ($Y_{i}^{a}$ is counterfactual when $A_{i} \neq a$). We thus rely on a deterministic potential outcome framework, where each level of treatment corresponds to only one outcome for each individual (but its value typically differs between individuals) \citep{Robins1989, VanderWeele2012}

We will consider only two treatment levels $\{0,1\}$ with $0$ indicating no treatment. Thus, the individual causal effect of an arbitrary individual $i$ is defined as $Y_{i}^{1}-Y_{i}^{0}$ \citep{Hernan2019}. When we discuss the random variable describing the heterogeneity of the potential outcomes in the population, we do not subscript the variables. We must make some identification assumptions to relate the distribution of potential outcomes to the distribution of observed outcomes. First of all, it is necessary to have access to a set of measured features $\boldsymbol{X}$ 
so that the treatment assignment is conditionally independent of the potential outcomes. \begin{assumption}{\textbf{Conditional exchangeability}}\label{CH2A1}
	$$A \independent Y^{0}, Y^{1}  \mid \boldsymbol{X}$$
\end{assumption}
\noindent This independence is called conditional exchangeability (or unconfoundedness) and implies the absence of unmeasured confounding that cannot be verified with observational data \citep{Hernan2019}. Then there are no features, other than $\boldsymbol{X}$, that $Y^0$ or $Y^1$ depend on and that differ in distribution between individuals with $A~{=}~1$ and $A~{=}~0$. Since we are interested in causal effect heterogeneity, the set of features $\boldsymbol{X}$ will also contain modifiers $\boldsymbol{X}_{\text{m}}$ (i.e.~$\exists \boldsymbol{x}_{1}, \boldsymbol{x}_{2}{:}~\mathbb{E}[Y^1-Y^0 \mid \boldsymbol{X}_{\text{m}}=\boldsymbol{x}_{1}] \neq \mathbb{E}[Y^1-Y^0 \mid \boldsymbol{X}_{\text{m}}=\boldsymbol{x}_{2}]$ \citep{VanderWeele2009}) next to the confounders that are necessary to obtain the independence. A feature can be only a modifier, only a confounder, or both, all on the additive scale. For a feature $L$ that is only a confounder but not a modifier $\forall l{:}~\mathbb{E}[Y^1-Y^0 \mid L=l, \boldsymbol{X}_{\text{m}}=\boldsymbol{x}] = \mathbb{E}[Y^1-Y^0 \mid \boldsymbol{X}_{\text{m}}=\boldsymbol{x}]$, where $\boldsymbol{X}_{\text{m}}$ represents the feature set $\boldsymbol{X}$ without $L$. 

Furthermore, we need to assume that the observed outcome of an individual equals the potential outcome for the assigned treatment, referred to as causal consistency \citep{Cole2009}.  
\begin{assumption}{\textbf{Causal consistency}}\label{CH2A2}
	$$Y_{i}~{=}~Y_{i}^{A_{i}}$$
\end{assumption} \noindent Causal consistency is also referred to as the stable unit treatment value assumption (SUTVA)\citep{Imbens2015}. Causal consistency implies that potential outcomes are independent of the treatment levels of other individuals (no interference) and that there are no different versions of the exposure levels. Causal consistency can also not be verified with data. 

Finally, the probability of receiving treatment should be bounded away from 0 and 1 for all levels of $\boldsymbol{X}$,  referred to as positivity \citep{Hernan2019}.
\begin{assumption}{\textbf{Positivity}}\label{CH2A3} $$%\exists \eta>0~\text{such that }
\forall \boldsymbol{x}{:}~ 0 < \mathbb{P}(A~{=}~1  \mid \boldsymbol{X}{=}\boldsymbol{x}) < 1 $$\end{assumption} \noindent Positivity is also known as overlap \citep{Imbens2015}.

As in \citet[Section 6]{Athey2019}, by causal consistency, we use the parameterization
\begin{equation}\label{eq:Y}
Y_{i} ~{=}~ Y^{0}_{i} + b_{i}A_{i}  , 
\end{equation} where $b_{i}$ is the ITE of individual $i$, so that $Y_{i}^{1}~{=}~Y^{0}_{i}+b_{i}$. The conditional mean of $b_{i}$ given features $\boldsymbol{X}_{i}$ equals the CATE $\tau(\boldsymbol{X}_{i})$, where $\tau(\boldsymbol{x}) = \mathbb{E}[Y^1-Y^0 \mid \boldsymbol{X}=\boldsymbol{x}]$. The ITE can thus be divided into $\tau(\boldsymbol{X}_{i})$ and the individual deviation from the CATE that is referred to as $U_{1i}$.  For our purposes, it helps to rewrite Equation \eqref{eq:Y} as
\begin{equation}\label{CH2eq:Y}
Y_{i} ~{=}~ \theta_{0}(\boldsymbol{X}_{i}) + N_{Yi} + \left(\tau(\boldsymbol{X}_{i})+U_{1i} \right)A_{i}, 
\end{equation} where $\theta_{0}(\boldsymbol{x})~{=}~\mathbb{E}[Y_{i}^{0} \mid \boldsymbol{X}_{i}{=}\boldsymbol{x}]$, $N_{Yi}$ represents the deviation of $Y_{i}^{0}$ from $\theta_{0}(\boldsymbol{X}_{i})$, $\tau(\boldsymbol{x})~{=}~\mathbb{E}[b_{i}  \mid \boldsymbol{X}_{i}{=}\boldsymbol{x}]$, $\mathbb{E}[N_{Yi}  \mid \boldsymbol{X}_{i}{=}\boldsymbol{x}]~{=}~0$ and $\mathbb{E}[U_{1i}  \mid \boldsymbol{X}_{i}{=}\boldsymbol{x}]~{=}~0$. In this parameterization, the individual $Y^{0}$ and effect $b$ have been rewritten as the sum of their conditional expectations and zero mean deviations from these expectations. Note that other characteristics (different from the mean) of the $N_{Y}  \mid \boldsymbol{X}{=}\boldsymbol{x}$ and $U_{1}  \mid \boldsymbol{X}{=}\boldsymbol{x}$ distributions can depend on the value of $\boldsymbol{x}$. Furthermore, $U_{1}$ and $N_{Y}$~~can be dependent. The latter can never be studied from data without making additional assumptions due to the fundamental problem of causal inference, i.e.~we cannot observe the pair $(Y^{0}, Y^{1})$.

\subsection{Case study and data simulation}\label{CH2sec:2.2}
\begin{figure}[H]
	\captionsetup{width=0.9\textwidth}
	\centering
	\begin{subfigure}{.45\textwidth}
		%\centering
		\resizebox{1\textwidth}{!}{\includegraphics{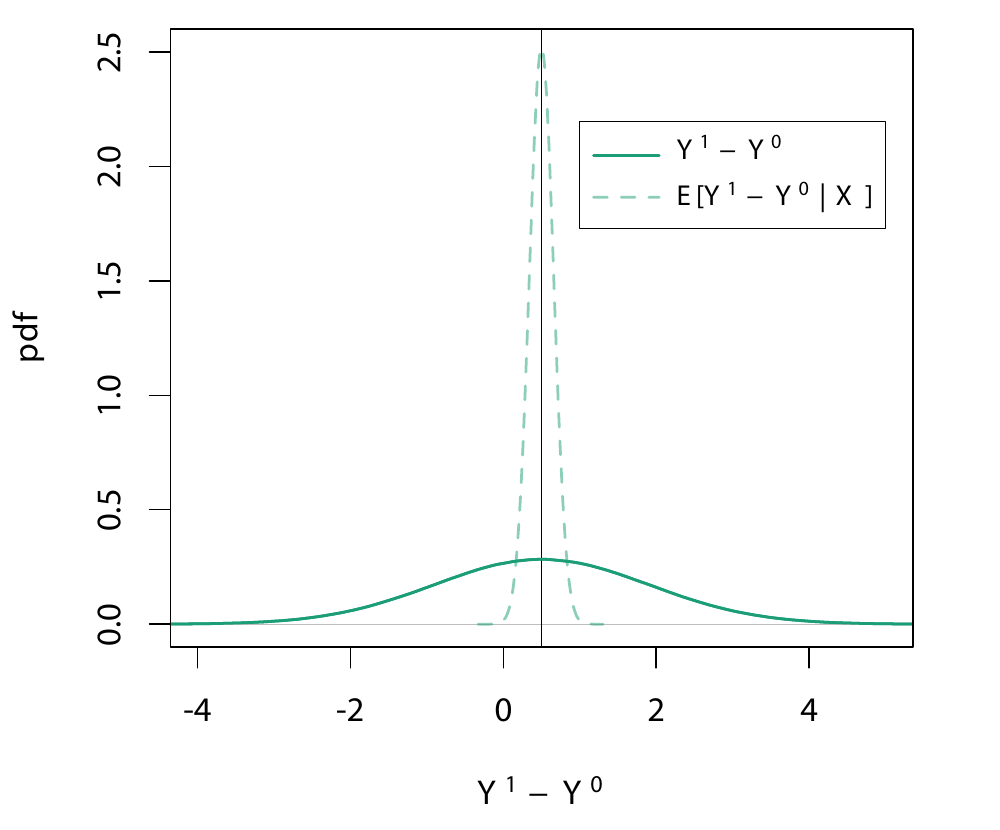}}
		\caption{}\label{CH21a}	
	\end{subfigure} 
	\begin{subfigure}{.45\textwidth}
		%\centering
		\resizebox{1\textwidth}{!}{\includegraphics{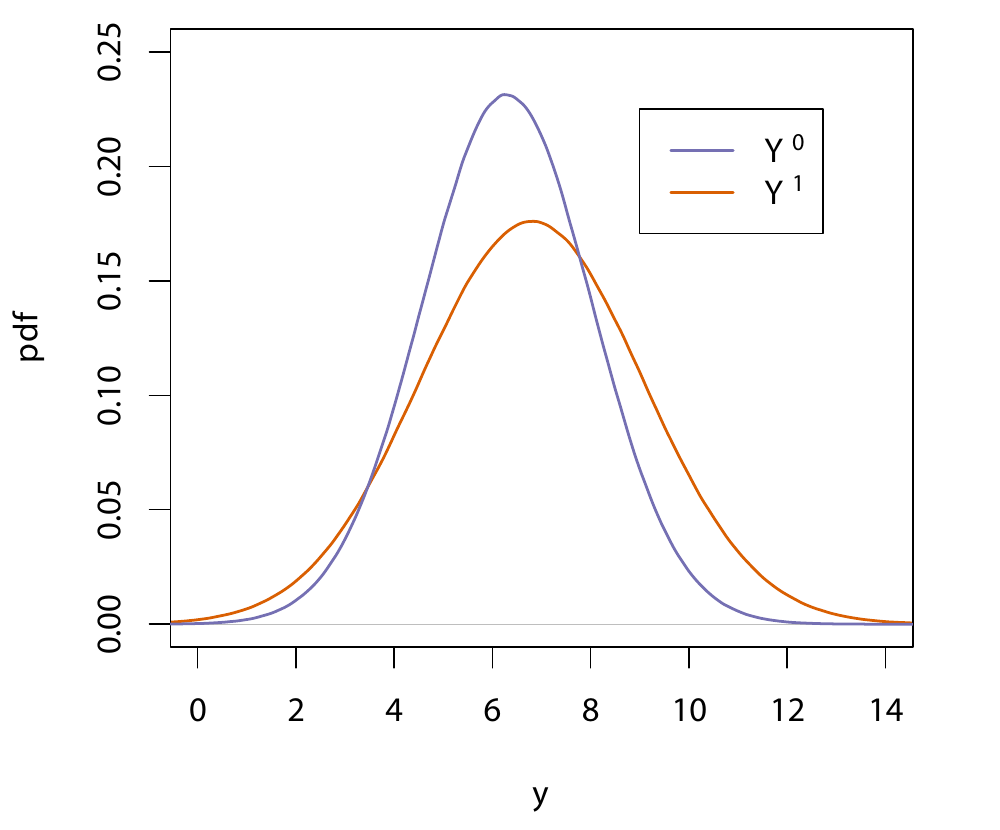}}
		\caption{}\label{CH21b}	
	\end{subfigure}
	
	\caption{The ITE (a, solid line) and potential outcome (b) distributions for the cause-effect relations underlying the simulation. The ATE is presented with a vertical solid black line (a). The distribution of a random CATE, thus for random $X_{\text{SBP}}$ and $X_{\text{sex}}$, is also presented (a, dotted line).}\label{CH2FIG1}
\end{figure}

To illustrate how the random conditional expectation $\mathbb{E}[Y^{1}-Y^{0} \mid \boldsymbol{X}]$ and  $Y^{1}-Y^{0}$ may differ in distribution, we simulate data based on the Framingham Heart Study (FHS) \citep{Mahmood2014}. We focus on the heterogeneity in the effect of non-alcoholic fatty liver disease on  a clinical precursor to heart failure, the left ventricular filling pressure \citep{Chiu2020}. The association found in the original work was adjusted for age, sex, smoking, alcohol use, diabetes, systolic blood pressure (SBP), antihypertensive-med use, lipid-lowering med use, total cholesterol, high-density lipoprotein cholesterol, triglycerides and fasting glucose. However, for this illustration, we assume that only sex (male $=0$ and female $=1$) and SBP are confounders. We will simulate the following cause-effect relations  
\begin{align}\label{eq:example}
A_{i}&~{=}~ \mathbbm{1}\left\{\frac{\exp(\alpha_{0}+\alpha_{\text{SBP}}X_{\text{SBP},i}+\alpha_{\text{sex}}X_{\text{sex},i})}{1+\exp(\alpha_{0}+\alpha_{\text{SBP}}X_{\text{SBP},i}+\alpha_{\text{sex}}X_{\text{sex},i})}>N_{Ai}\right\}\\ \nonumber
Y^{0}_{i}&~{=}~ \beta_{0} + \beta_{\text{sex}}X_{\text{sex},i} +\beta_{\text{SBP}}X_{\text{SBP},i} + N_{Yi}\\ \nonumber
Y^{1}_{i}&~{=}~ Y^{0}_{i} + \left(\tau_{0} + \tau_{\text{sex}}X_{\text{sex},i} +\tau_{\text{SBP}}X_{\text{SBP},i} +U_{1i}\right),
\end{align} where $X_{\text{sex},i}\sim \text{Ber}(p)$, $X_{\text{SBP},i}\sim \mathcal{N}(0, 1)$, $U_{1i}\sim \mathcal{N}(0,\sigma_{1}^{2})$, $N_{Yi} \sim \mathcal{N}(0, \sigma_{0}^{2})$, $N_{Ai} \sim \text{Uni}[0,1]$, and $U_{1i} \independent N_{Yi}$. %(and thus $U_{1i} \independent Y^{0}_{i}$). 
Moreover, there is no unmeasured confounding, i.e.~$N_{Ai} \independent N_{Yi}, U_{1i}$ so that $A_{i} \independent (Y^{1}_{i}, Y^{0}_{i})  \mid X_{\text{sex},i}, X_{\text{SBP},i}$.  By causal consistency, the observed outcome $Y_{i}~{=}~Y_{i}^{A}$ equals $Y_{i}^{1}$ when $A_{i}~{=}~1$ and $Y_{i}^{0}$ when $A_{i}~{=}~0$. The parameter values are obtained by fitting a linear mixed model for the relation of fatty liver disease and the left ventricular filling pressure adjusted for standardized SBP and sex, \begin{equation} Y ~{=}~ \beta_{0} + \beta_{\text{sex}}X_{\text{sex},i} +\beta_{\text{SBP}}X_{\text{SBP},i} + N_{Yi} + \left(\tau_{0} + \tau_{\text{sex}}X_{\text{sex},i} +\tau_{\text{SBP}}X_{\text{SBP},i} +U_{1i}\right)A_{i},\end{equation} to the subset of the FHS participants $(n=2356)$ as used by \citet{Chiu2020}. The values obtained with  \texttt{PROC LOGISTIC} and \texttt{PROC MIXED} in \texttt{SAS} equal $\alpha_{0}=-1.7$, $\alpha_{\text{sex}}=-0.1$, $\alpha_{\text{SBP}}=0.4$ (so that $\mathbb{P}(A~{=}~1  \mid X_{\text{sex}}=1)~{=}~0.15$ and $\mathbb{P}(A~{=}~1  \mid X_{\text{sex}}=0)~{=}~0.16$), $\beta_{0}=5.9$, $\beta_{\text{sex}}=0.8$,  $\beta_{\text{SBP}}=0.5$, $\tau_{0}=0.45$, $\tau_{\text{sex}}=0.1$, $\tau_{\text{SBP}}=0.15$,  $\sigma_{0}^{2}=1.6^2$ and $\sigma_{1}^2=1.4^2$. The distribution of $Y^{1}-Y^{0}$ %is a two-component (as a result of the modification by sex) Gaussian mixture, as illustrated 
is shown in Figure \ref{CH21a}, where $\mathbb{E}[Y^{1}-Y^{0}]~{=}~0.5$, $\sqrt{\text{var}(Y^{1}-Y^{0})}~{=}~1.41$ and $\mathbb{P}(Y^{1}-Y^{0}>0)~{=}~0.64$. Furthermore, the distribution of the conditional expectation $\mathbb{E}[Y^{1}-Y^{0} \mid X_{\text{SBP}}, X_{\text{sex}}]$ is shown  in Figure \ref{CH21a} with a standard deviation equal to $0.16$ and $\mathbb{P}\left( \mathbb{E}[ Y^{1} - Y^{0}  \mid X_{\text{SBP}}, X_{\text{sex}} ] > 0 \right) ~{=}~ 1.00$. The conditional expectation distribution seriously differs from that of the ITE due to the unmeasured (remaining) effect heterogeneity $(U_{1})$. For completeness, the distributions of $Y^1$ and $Y^0$ are presented in Figure \ref{CH21b}. Moreover, we simulate $X_{0}$, which is a measured variable associated with the level of the individual modifier $U_{1}$, $(U_{1}, X_{0})^{T} \sim \mathcal{N}\left(\boldsymbol{0}, \begin{pmatrix} \sigma_{1}^{2} & \rho \delta \sigma_{1}^{2}\\ \rho \delta \sigma_{1}^{2} & \delta^2 \sigma_{1}^{2} \end{pmatrix} \right)$. For $\rho>0$, $X_{0}$ is another measured modifier. Varying $\rho$ can thus be used to investigate cases where more of the latent individual effect modification can be explained while preserving the  distribution of $Y^{1}-Y^{0}$. All programming codes used for this work can be found online at \url{https://github.com/RAJP93/CATE}.

\subsection{Causal random forest}\label{CH2sec:2.1}
Since the actual causal effects are not observed, defining an appropriate loss function is not straightforward, so using ML methods to study causal effect heterogeneity is challenging \citep{Athey2016}. Causal trees \citep{Athey2016} and causal forests \citep{Wager2018} have been introduced to draw inferences in case of complex causal effect heterogeneity. These papers mainly focus on data from randomized experiments and suggest adding traditional propensity score methods \citep{Athey2016} or using a different algorithm less sensitive to the complexity of the treatment effect function \citep{Wager2018} for observational studies. The CRF procedure as implemented in the \texttt{causal\_forest} function from the R-package \texttt{grf} is an example of a generalized random forest (GRF) \citep{Athey2019}. This CRF can be used on data from randomized experiments and observational studies without unmeasured confounding. This ML method is a random forest-based variant of the R-learner \citep{Nie2020}. As mentioned in the introduction, R-learners focus on the relation between normalized outcome and treatment assignment as originally used by \citet{Robinson1988} 
%\begin{equation}\label{eq:R}
%Y_{i}  \mid \boldsymbol{X}
%_{i}{{=}}\boldsymbol{x} - \mathbb{E}[Y_{i}  \mid \boldsymbol{X}
%_{i}{{=}}\boldsymbol{x}] ~{=}~ \left(A_{i} - \mathbb{E}[A_{i}  \mid %\boldsymbol{X}
%_{i}{{=}}\boldsymbol{x}]\right) \widetilde{{}\tau}(\boldsymbol{x}) + \widetilde{{}N}_{Yi},\end{equation} 
\begin{equation}\label{eq:R}
Y_{i} - m(\boldsymbol{X}_{i}) = \left(A_{i} - e(\boldsymbol{X}_{i})\right)
 \widetilde{{}\tau}(\boldsymbol{X}_{i}) + \widetilde{{}N}_{Yi},\end{equation}
where $m(\boldsymbol{x})=\mathbb{E}[Y\mid \boldsymbol{X}{=}\boldsymbol{x}]$,  $e(\boldsymbol{x})=\mathbb{E}[A\mid \boldsymbol{X}{=}\boldsymbol{x}]$ and $\forall \boldsymbol{x}, a{:}~\mathbb{E}\left[\widetilde{{}N}_{Yi}~{\mid}~A_{i}{{=}}a, \boldsymbol{X}
_{i}{{=}}\boldsymbol{x}\right]=0$. Associational quantities are presented with a tilde, and causal quantities are without. Using this convention, $\widetilde{{}\tau}(\boldsymbol{x})$ represents the conditional association measure of $A$ and $Y$ given $\boldsymbol{X}{=}\boldsymbol{x}$. This association will equal the CATE for  $\boldsymbol{X}{=}\boldsymbol{x}$, $\tau(\boldsymbol{x})$, under certain identification assumptions, as we will show next.  Using parameterization \eqref{CH2eq:Y} one can derive, for $\boldsymbol{X}_{i}=\boldsymbol{x}$, 
\begin{align}
Y_{i} - m(\boldsymbol{x}) &= \left(\theta_{0}(\boldsymbol{x}) + N_{Yi} + \left(\tau(\boldsymbol{x})+U_{1i} \right)A_{i} \right)- \left(\theta_{0}(\boldsymbol{x})  + \mathbb{E}\left[ \left(\tau(\boldsymbol{x})+U_{1i}\right)A_{i} \mid \boldsymbol{X}_{i}{=}\boldsymbol{x}\right]\right)\\ \nonumber
&~{=}~ \left(A_{i}-e(\boldsymbol{x})\right)\tau(\boldsymbol{x}) + \left(A_{i}U_{1i} - \mathbb{E}\left[A_{i} U_{1i}  \mid \boldsymbol{X}_{i}{=}\boldsymbol{x}\right] + N_{Yi}\right), \nonumber
\end{align} where the distribution of $A_{i}, U_{1i}$ and $N_{Yi}$ can depend on the value of $\boldsymbol{x}$. If $\mathbb{E}[U_{1}+N_{Y}~{\mid}~A{{=}}1, \boldsymbol{X}{=}\boldsymbol{x}] \neq \mathbb{E}[N_{Y}~{\mid}~A{=}0, \boldsymbol{X}{=}\boldsymbol{x}]$, then $\widetilde{{}\tau}(\boldsymbol{x}) \neq \tau(\boldsymbol{x})$. This can occur when there is remaining (unmeasured) confounding after adjusting for $\boldsymbol{X}$. However, in absence of unmeasured confounding, {i.e.}~$A \independent N_{Y}, U_{1}$, for $\boldsymbol{X}_{i}=\boldsymbol{x}$,
\begin{equation*}
Y_{i} - m(\boldsymbol{x}) =  \left(A_{i}-e(\boldsymbol{x}) \right)\tau(\boldsymbol{x}) + \left(A_{i}U_{1i} + N_{Yi}\right),
\end{equation*} where $\forall \boldsymbol{x}{:}~\mathbb{E}[U_{1i}+N_{Yi}~{\mid}~A_{i}{=}1, \boldsymbol{X}
_{i}{=}\boldsymbol{x}]{=}\mathbb{E}[N_{Yi}~{\mid}~A_{i}{=}0 , \boldsymbol{X}
_{i}{=}\boldsymbol{x}]{=}0$. Then, $\widetilde{{}\tau}(\boldsymbol{x})$ equals $\tau(\boldsymbol{x})$, the CATE for $\boldsymbol{X}{=}\boldsymbol{x}$.  

In the absence of unmeasured confounding, the R-learner thus allows us to estimate (or predict) the CATEs, $\tau(\boldsymbol{x})$, from observational data. The GRF implementation of the CRF starts by predicting $m(\boldsymbol{x}_{i})$ and $e(\boldsymbol{x}_{i})$ for each $i$ by fitting two separate regression forests consisting of honest trees (each tree is fitted on a random subsample of half the sample size) \citep{Wager2018}. The out-of-bag predictions $\hat{{}m}^{-i}(\boldsymbol{x}_{i})$ and $\hat{{}e}^{-i}(\boldsymbol{x}_{i})$ are only based on those trees that did not use individual $i$ for training, and used to create the centered outcomes $\widetilde{{}Y}_{i}~{=}~Y_{i}-\hat{{}m}^{-i}(\boldsymbol{x}_{i})$ and $\widetilde{{}A}_{i}~{=}~A_{i}-\hat{{}e}^{-i}(\boldsymbol{x}_{i})$ for individual $i$. %Out-of-bag  predictions $\hat{{}m}^{-i}(\boldsymbol{x}_{i})$ and $\hat{{}e}^{-i}(\boldsymbol{x}_{i})$ refer to estimates only based on trees that did not use observation $i$ during training.
Subsequently, for a new set of features $\boldsymbol{x}$, similarity weights $\alpha_{j}(\boldsymbol{x})$ are produced for each observation in the sample, and $\widetilde{{}\tau}(\boldsymbol{x})$ (that can be a complex function of $\boldsymbol{x}$) is estimated as
\begin{equation}\label{CH2eq:CATE} \hat{{}\widetilde{{}\tau}}(\boldsymbol{x}) ~{=}~ \frac{\sum_{i{=}1}^{n} \alpha_{i}(\boldsymbol{x}) \left( Y_{i} - \hat{{}m}^{-i}(\boldsymbol{x}
	_{i})\right)\left(A_{i} - \hat{{}e}^{-i}(\boldsymbol{x}
	_{i})\right)}{\sum_{i{=}1}^{n} \alpha_{i}(\boldsymbol{x}) \left(A_{i} - \hat{{}e}^{-i}(\boldsymbol{x}
	_{i})\right)},  
\end{equation} see \citet{Athey2019b} for more details. The $\alpha_{j}(\boldsymbol{x})$ are obtained by first growing a set of $B$ (user-specified, default is $2000$) trees for $\widetilde{Y}$. For each tree, a random subsample $\mathcal{I}$ of the available data is taken (fraction is user-specified, default is $0.5$). The subsample is randomly divided into (by default) equally sized $\mathcal{J}_{1}$ and $\mathcal{J}_{2}$. The honest decision tree, in the sense of \citet{Wager2018}, is only fitted on $\mathcal{J}_{1}$ and optimizes the heterogeneity in the effect of $\widetilde{{}A}$ on $\widetilde{{}Y}$~ between the different nodes using gradient-based approximations of treatment-effect estimates in candidate children notes, see \citet[Section 2.3]{Athey2019}. The similarity weights, $\alpha_{bj}$, are first estimated per tree $b$ and are non-zero (and equal) for those elements of $\mathcal{J}_{2}$ that fall in the same leaf as $\boldsymbol{x}$, and are averaged over all trees to obtain $\alpha_{j}$. For individuals from the original dataset, (out-of-bag) predictions are made by averaging the similarity weights only over trees that did not use this particular observation during training. 
The ATE is estimated using the augmented inverse probability weighting (AIPW) estimator \citep{Robins1995} and equals
\begin{equation}\label{CH2eq:ATE} \frac{1}{n} \sum_{i{=}1}^{n} \left( \hat{{}\widetilde{{}\tau}}(\boldsymbol{x}
_{i}) + A_{i} \frac{Y_{i}-\hat{{}\mu}_{1i}(\boldsymbol{x}
	_{i})}{\hat{{}e}^{-i}(\boldsymbol{x}_{i})} - (1-A_{i}) \frac{Y_{i}-\hat{{}\mu}_{0i}(\boldsymbol{x}_{i})}{1-\hat{{}e}^{-i}(\boldsymbol{x}_{i})} \right),\end{equation} where
$\hat{{}\mu}_{1i}(\boldsymbol{x}
_{i})~{=}~ \hat{{}m}^{-i}(\boldsymbol{x}
_{i})+(1-\hat{{}e}^{-i}(\boldsymbol{x}_{i})) \hat{{}\widetilde{{}\tau}}(\boldsymbol{x}
_{i})$ and $\hat{{}\mu}_{0i}(\boldsymbol{x}
_{i})~{=}~ \hat{{}m}^{-i}(\boldsymbol{x}
_{i})-\hat{{}e}^{-i}(\boldsymbol{x}
_{i}) \hat{{}\widetilde{{}\tau}}(\boldsymbol{x}
_{i})$. The AIPW estimator is double robust, i.e.~consistent if $ \hat{{}\widetilde{{}\tau}}(\boldsymbol{x})$ or $\hat{{}e}(\boldsymbol{x})$ is a consistent estimator of $\tau(\boldsymbol{x})$ or $e(\boldsymbol{x})$ respectively. 

In this work, we fit a CRF to the simulated data as described in Section \ref{CH2sec:2.2} to estimate the $(X_{\text{sex}}, X_{\text{SBP}}, X_{0})$-CATE for each individual. We vary the sample size, $n \in \{200, 2000, 20000 \}$, and the correlation between the unmeasured modifier $U_{1}$ and the measured $X_{0}$, $\rho \in \{0, 0.25, 0.5, 0.75, 1 \}$, while fixing $\delta=2$. We use the default settings of the \texttt{causal\_forest} function, except for the $n=200$ settings where we set \texttt{min.node.size{=}1}. For each simulation, we compute the empirical standard deviation (SD) and positive effect probability (PEP), $\mathbb{P}(Y^1-Y^0~{>}~0)$, of the estimated CATE distribution to estimate the SD and PEP of the ITE distribution, respectively. As presented in Section \ref{CH2sec:2.2}, their actual values equal $1.41$, and $0.64$, respectively. The ATE (equal to $0.5$) is estimated with the AIPW estimator using the \texttt{average\_treatment\_effect} function of the \texttt{grf} package. Furthermore, based on $1000$ bootstrap samples, we estimate $95\%$ confidence intervals (CIs) for all three characteristics. Based on $1000$ simulations, we estimate the bias, mean squared error (MSE) and coverage for the different settings. Finally, we estimate the ITE distribution per simulation with a Gaussian kernel density estimator over the estimated CATEs using the $\texttt{density}$ function in R with the default settings.

\section{Results}\label{CH2CH2:sec3}
The bias, MSE and coverage for the ATE, SD and PEP of the ITE distribution based on the CATE distribution, estimated with the CRF, are presented in Table \ref{CH2tab:1} for the different settings. 
%TC:ignore
\begin{table}[h!]
	\captionsetup{width=0.9\textwidth}
	\caption{Bias, mean squared error (MSE) and coverage of the estimated ATE, SD and PEP ($\mathbb{P}(Y^1-Y^0>0)$) of the ITE distribution using the characteristics of the CATE distribution distribution as an estimator, based on $1000$ simulated samples per setting. }\label{CH2tab:1}
	\centering
	%\resizebox{0.8\textwidth}{!}{
	\begin{tabular}{cc||ccc|ccc|ccc}
		\multicolumn{2}{c||}{} & \multicolumn{3}{c|}{Bias} & \multicolumn{3}{c|}{MSE} & \multicolumn{3}{c}{Coverage} \\ \hline
		$\rho$    & $n$      & ATE     & SD     & PEP   & ATE    & SD     & PEP   & ATE      & SD      & PEP     \\ \hline
	0 & 200 & 0.05 & -0.97 & 0.18 & 0.20 & 0.97 & 0.08 & 0.95 & 0.02 & 0.83 \\ 
  0 & 2000 & -0.00 & -1.18 & 0.33 & 0.02 & 1.39 & 0.11 & 0.94 & 0.00 & 0.20 \\ 
  0 & 20000 & 0.00 & -1.18 & 0.35 & 0.00 & 1.40 & 0.12 & 0.95 & 0.00 & 0.00 \\ 
  0.25 & 200 & 0.01 & -0.94 & 0.16 & 0.20 & 0.91 & 0.07 & 0.94 & 0.03 & 0.86 \\ 
  0.25 & 2000 & 0.00 & -1.07 & 0.28 & 0.02 & 1.15 & 0.08 & 0.94 & 0.00 & 0.29 \\ 
  0.25 & 20000 & -0.00 & -1.02 & 0.26 & 0.00 & 1.04 & 0.07 & 0.95 & 0.00 & 0.00 \\ 
  0.50 & 200 & 0.02 & -0.87 & 0.14 & 0.19 & 0.80 & 0.06 & 0.95 & 0.07 & 0.86 \\ 
  0.50 & 2000 & 0.01 & -0.82 & 0.15 & 0.02 & 0.69 & 0.03 & 0.95 & 0.00 & 0.56 \\ 
  0.50 & 20000 & 0.00 & -0.71 & 0.12 & 0.00 & 0.51 & 0.02 & 0.94 & 0.00 & 0.01 \\ 
  0.75 & 200 & 0.04 & -0.73 & 0.12 & 0.16 & 0.58 & 0.04 & 0.95 & 0.19 & 0.87 \\ 
  0.75 & 2000 & 0.01 & -0.53 & 0.06 & 0.01 & 0.29 & 0.01 & 0.95 & 0.00 & 0.85 \\ 
  0.75 & 20000 & 0.00 & -0.38 & 0.05 & 0.00 & 0.15 & 0.00 & 0.94 & 0.00 & 0.51 \\ 
  1.00 & 200 & 0.04 & -0.56 & 0.09 & 0.14 & 0.37 & 0.03 & 0.95 & 0.37 & 0.88 \\ 
  1.00 & 2000 & 0.01 & -0.20 & 0.01 & 0.01 & 0.05 & 0.00 & 0.96 & 0.42 & 0.93 \\ 
  1.00 & 20000 & 0.00 & -0.04 & 0.00 & 0.00 & 0.00 & 0.00 & 0.95 & 0.76 & 0.94 \\ 
	\end{tabular}%
	
\end{table}
%TC:endignore

\begin{figure}[H]
	\centering
	\resizebox{0.9\linewidth}{!}{\includegraphics{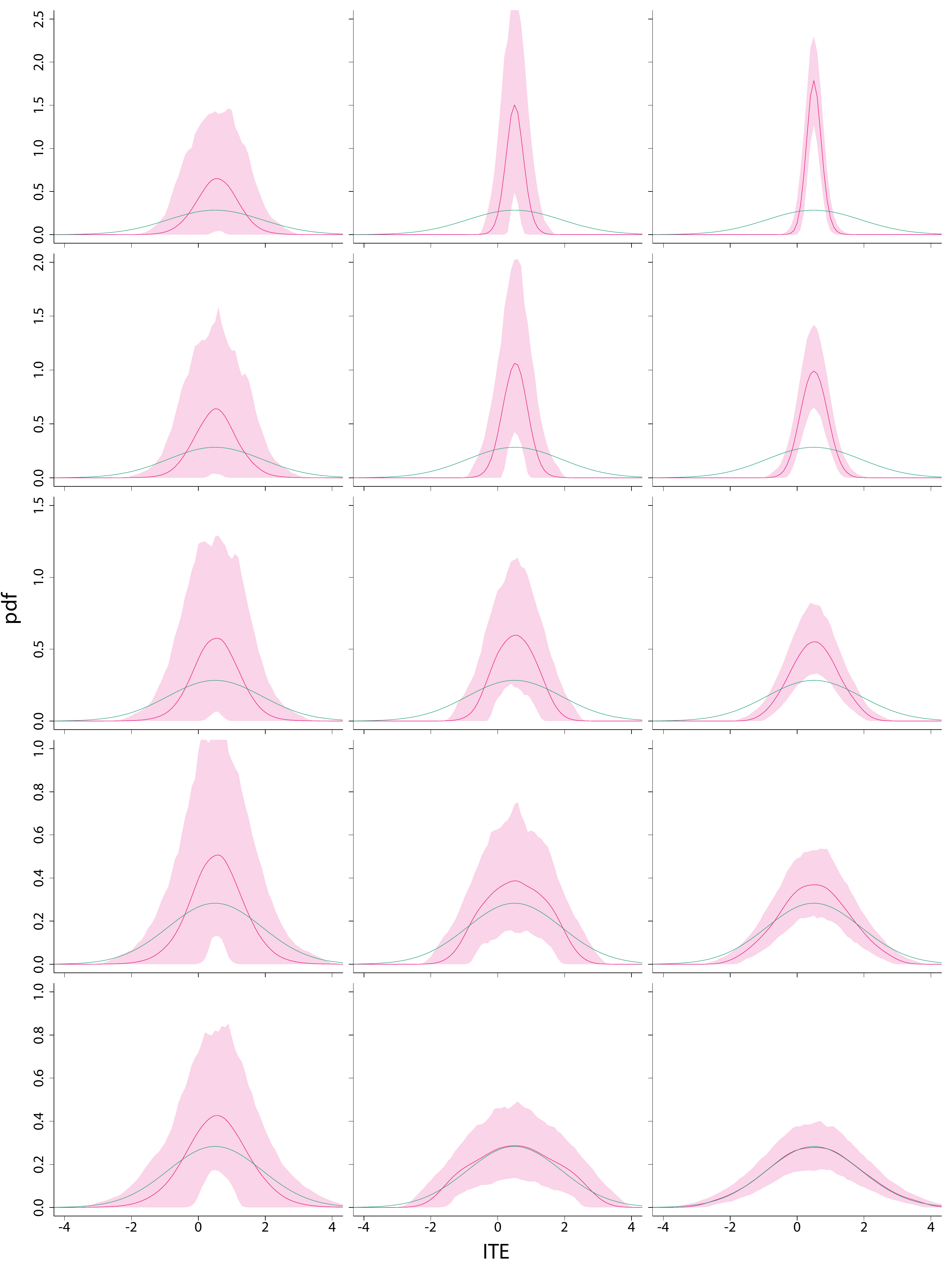}}
	\caption{Pointwise empirical mean (opaque pink) as well as the $2.5\%$ and $97.5\%$ quantiles (transparent pink) of the estimated ITE densities of $1000$ simulations. Sample size per simulation varies over the columns and equals $200$ (left), $2000$ (middle) and $20000$ (right). The association of the measured $X_{0}$ and unmeasured modifier $U_{1}$ varies per row equal to $0, 0.25, 0.50, 0.75$ and $1$ (top to bottom). Furthermore, the actual ITE distribution is presented (green). }\label{CH2fig2}
\end{figure}

In the absence of features ($X_{0}$) that are associated with the unmeasured modifier $U_{1}$, i.e.~when $\rho~{=}~0$, the variability in the ITE is seriously underestimated when using the CATE distribution as a proxy as shown in the first row of Figure \ref{CH2fig2}. Therefore, the SD and PEP of the distribution of the conditional expectation ($\mathbb{E}[Y^1-Y^0  \mid \boldsymbol{X}]$) are biased estimators of the characteristics of the ITE distribution. The bias is the lowest for a small sample size due to a finite-sample effect for both the SD and PEP. For $n=200$, the coverage of the PEP is not much off. For larger sample sizes, the CATE distribution can be estimated more precisely. Then, the bias increases, and the coverage decreases. 

We observe the same trend for $\rho~{=}~0.25$. However, for $\rho\geq 0.50$, the bias is more extensive for small sample sizes. In the latter cases, the finite-sample effect of the CATE distribution estimator no longer compensates for the difference between the ITE and CATE distribution. Nevertheless, the uncertainty in the estimate for small sample sizes still results in higher coverage of the PEP. For $\rho~{=}~0.75$, the CATE distribution becomes a reasonable proxy for the ITE distribution, as seen from the fourth row in Figure \ref{CH2fig2}. Finally, for $\rho~{=}~1$, $U_{1}$ equals $X_{0}$, and there is thus no unmeasured effect modification. In this case, the flexible ML estimation of the CATE distribution can be used to estimate the ITE distribution and understand the variability in the treatment effect. Indeed, the bias of the SD and the PEP become small,  and the coverage approaches the nominal probability. The bias of the SD is still not neglectable, so the coverage of the SD deviates from the nominal probability. 

\newpage

\section{From conditional means to conditional distributions}\label{CH2CH2:sec4}
We presented examples in which the distribution of $\mathbb{E}[Y^{1}-Y^{0}  \mid \boldsymbol{X}]$ differs from that of $Y^{1}-Y^{0}$. To overcome this issue, we should consider the remaining effect heterogeneity.  In this section, we show that the variance of $Y^{1}-Y^{0}  \mid \boldsymbol{X}{=}\boldsymbol{x}$ is only identifiable when we are willing to make another causal assumption. Under this assumption, we can extend the CRF algorithm by also estimating the conditional variance of the effect for each individual. For the parameterization in Equation \eqref{CH2eq:Y}, the variance of $Y^{1}-Y^{0}  \mid \boldsymbol{X}{=}\boldsymbol{x}$ equals $\sigma_{1}^{2}(\boldsymbol{x})=\mathbb{E}[(U_{1})^{2}  \mid \boldsymbol{X}{=}\boldsymbol{x}]$. As derived in Appendix \ref{CH2app:Y2RC}, in the absence of unmeasured confounding, the Robinson decomposition of the squared observations enables us to estimate $$\Delta(\boldsymbol{x}) = \tau(\boldsymbol{x})^{2} + \sigma_{1}^{2}(\boldsymbol{x})  + 2\tau(\boldsymbol{x})\theta_{0}(\boldsymbol{x})  + 2\mathbb{E}[N_{Y} U_{1} \mid \boldsymbol{X}{=}\boldsymbol{x}] $$ from observational data. Subsequently, via estimation of $\tau(\boldsymbol{x})$ and $\theta_{0}(\boldsymbol{x})$, $$\widetilde{\sigma_{1}}^{2}(\boldsymbol{x})= \sigma_{1}^{2}(\boldsymbol{x}) + 2\mathbb{E}[N_{Y} U_{1}  \mid \boldsymbol{X}{=}\boldsymbol{x}]$$ can be estimated by substraction. Since $\mathbb{E}[U_{1}\mid\boldsymbol{X}=\boldsymbol{x}]$ and  $\mathbb{E}[N_{Y}\mid\boldsymbol{X}=\boldsymbol{x}]$ equal 0, $\widetilde{\sigma_{1}}^{2}(\boldsymbol{x})$ represents the sum of the conditional variance of the ITE and twice the conditional covariance of $Y^{0}$ and the ITE. However, as a result of the fundamental problem of causal inference, %the joint distribution of $(N_{Y}, U_{1}) \mid \boldsymbol{X}{=}\boldsymbol{x}$, and neither 
$\mathbb{E}[N_{Y} U_{1}  \mid \boldsymbol{X}{=}\boldsymbol{x}]$, the conditional expectation of the product of the deviation of $Y^{0}$ from $\theta(\boldsymbol{x})$ and the deviation of $Y^{1}-Y^{0}$ from $\tau(\boldsymbol{x})$, is not identifiable. So, we cannot estimate %the variance of $Y^{1}-Y^{0}  \mid \boldsymbol{X}{=}\boldsymbol{x}$, 
$\sigma_{1}^{2}(\boldsymbol{x})$ %=\mathbb{E}[(U_{1})^{2}  \mid \boldsymbol{X}{=}\boldsymbol{x}]$
 without an additional (cross-world) assumption. 

If we can assume that $U_{1} \independent N_{Y}  \mid \boldsymbol{X}{=}\boldsymbol{x}$, then $\mathbb{E}[N_{Y} U_{1}  \mid \boldsymbol{X}{=}\boldsymbol{x}]~{=}~0$ and $\widetilde{\sigma_{1}}^{2}(\boldsymbol{x})=\sigma_{1}^{2}(\boldsymbol{x})$, so that %or equivalently  $U_{1} \independent Y^{0}  \mid \boldsymbol{X}{=}\boldsymbol{x}$, 
the variance of the causal effect given $\boldsymbol{X}$ becomes identifiable. %as Equation \eqref{CH2eq:varx} 
The assumption implies conditional independence of the outcome under no treatment and the effect, i.e.~the deviation of $Y^0 \mid \boldsymbol{X}{=}\boldsymbol{x}$ from $\mathbb{E}[Y^0 \mid \boldsymbol{X}{=}\boldsymbol{x}]$ is independent of the deviation of $Y^1-Y^0  \mid \boldsymbol{X}{=}\boldsymbol{x}$ from the CATE. Therefore, we refer to this assumption as conditional independent effect deviation.  \begin{assumption}{\textbf{Conditional independent effect deviation}}\label{CH2A4}
	$$ Y^{1}-Y^{0} \independent Y^{0}  \mid \boldsymbol{X}{=}\boldsymbol{x} $$\end{assumption} \noindent Assumption \ref{CH2A4} implies that all features that affect both $Y^0$ and $Y^1-Y^0$ should be contained in $\boldsymbol{X}$.  As an example, one could think of the effectiveness of  medical drugs that depends on the amount of enzyme present for an individual, while the presence of the enzyme itself does not inform on the outcome of interest in the absence of the drug. The antiplatelet medicine Clopidogrel reduces the risk of stroke and myocardial infarction in individuals with acute coronary syndrome, but its effect depends on its conversion to an active metabolite which is accomplished by the cytochrome P450 2C19 (CYP2C19) enzyme \citep{Craig2022}. For individuals with a CYP2C19 gene mutation, the drug is known to have a reduced antiplatelet effect; the CYP219 gene thus results in effect heterogeneity. However, there is no reason to believe that the phenotype affects platelet aggregation in the absence of the drug. In cases where $Y^0$ is still expected to inform on the value of $Y^1-Y^0$ given the levels of $\boldsymbol{X}$, the identification assumption does not apply. Similar to Assumption \ref{CH2A1}, this causal assumption cannot be verified with data as it concerns unmeasured features that affect both $Y^{0}$ and $Y^{1}-Y^{0}$ and should be judged by experts in the field of application. However, in contrast to Assumption \ref{CH2A1}, no reason guarantees that Assumption \ref{CH2A4} holds in a randomized experiment.
 
Also, in the case where sufficient features are measured so that the ITE equals the CATE, Assumption \ref{CH2A4} holds as $\forall i{:}~ U_{1i}~{=}~0$ and thus independent of $Y_{i}^{0}$. %so that $\mathbb{E}[ (U_{1})^{2} \mid \boldsymbol{X}{=}\boldsymbol{x}] + \mathbb{E}[N_{Y} U_{1}  \mid \boldsymbol{X}{=}\boldsymbol{x}] ~{=}~ 0$ anyway. 

 If Assumption \ref{CH2A4} (in addition to \ref{CH2A1}, \ref{CH2A2} and \ref{CH2A3}) holds, then $\text{var}(Y^1-Y^0  \mid \boldsymbol{X}{=}\boldsymbol{x})$ can be estimated %as \eqref{CH2eq:varx} 
 with an extended CRF as presented in Algorithm \ref{alg:EGRF}. %well explain next. 

%TC:ignore

\begin{minipage}{0.9\linewidth}
\centering
\begin{algorithm}[H]
\caption{Extended CRF}\label{alg:EGRF}
\begin{algorithmic}[1]
\State{\textbf{input} features $\boldsymbol{x}$}
\State{Run the traditional CRF}
 \begin{algsubstates}
\State $\phantom{\text{iiii}}\mathllap{~}$ Save predicted CATE, $\hat{{}\widetilde{{}\tau}}(\boldsymbol{x})$
\State $\phantom{\text{iii}}\mathllap{~}$  $\forall i{:}~$save the predicted $A_{i}$, $\hat{{}e}^{-i}(\boldsymbol{x}_{i})$
\State $\phantom{\text{ii}}\mathllap{~}$ Save the predicted $Y^{0}$, $\hat{{}\widetilde{\theta_{0}}}(\boldsymbol{x})$ \Comment{\parbox[t]{.5\linewidth}}{When $\exists i{:}~\boldsymbol{x}_{i}{=}\boldsymbol{x}$ then  $\hat{{}\widetilde{\theta_{0}}}(\boldsymbol{x})$ equals $\hat{{}\mu}_{0i}(\boldsymbol{x}_{i})$ as defined after Equation \eqref{CH2eq:ATE} by only considering those trees that did not use individual $i$ for training.\\}
\State $\phantom{\text{ii}}\mathllap{~}$ Save the similarity weights $\alpha_{i}(\boldsymbol{x})$ \Comment{\parbox[t]{.5\linewidth}}{When $\exists i{:}~\boldsymbol{x}_{i}{=}\boldsymbol{x}$ then  $\forall j{\neq}i{:}~\alpha_{j}(\boldsymbol{x}_{i})$ are created by averaging similarity weights over trees where $i \not \in \mathcal{J}_{1}$.\\}
 \end{algsubstates}
 \State Fit a separate regression forest to estimate $h(\boldsymbol{x}_{i})=\mathbb{E}[(Y_{i})^{2}  \mid \boldsymbol{X}_{i}{=}\boldsymbol{x}_{i}]$
 \begin{algsubstates}
\State $\forall i{:}~$ obtain the out of bag predictions of $(Y_{i})^{2}$, $\hat{{}h}^{-i}(\boldsymbol{x}_{i})$
 \end{algsubstates}
 \NoNumber{~}
 
\State Estimate $\mathbb{E}[(Y^{1})^{2}-(Y^{0})^{2} \mid \boldsymbol{X}{=}\boldsymbol{x}]$ as $$\hat{{}\widetilde{\Delta}}(\boldsymbol{x}) = \frac{\sum_{i{=}1}^{n} \alpha_{i}(\boldsymbol{x}) \left( (Y_{i})^{2} - \hat{{}h}^{-i}(\boldsymbol{x}_{i})\right)\left(A_{i} - \hat{{}e}^{-i}(\boldsymbol{x}_{i})\right)}{\sum_{i{=}1}^{n} \alpha_{i}(\boldsymbol{x}) \left(A_{i} - \hat{{}e}^{-i}(\boldsymbol{x}_{i})\right)}.$$
\State Estimate $\mathbb{E}[ (U_{1})^{2} \mid \boldsymbol{X}{=}\boldsymbol{x}] + 2\mathbb{E}[N_{Y} U_{1}  \mid \boldsymbol{X}{=}\boldsymbol{x}]$ as 
$$ {\hat{{}\widetilde{\sigma_{1}}}}^{2}(\boldsymbol{x}) = \hat{{}\widetilde{\Delta}}(\boldsymbol{x}) -  \hat{{}\widetilde{{}\tau}}(\boldsymbol{x})^{2} - 2\hat{{}\widetilde{{}\tau}}(\boldsymbol{x})\hat{{}\widetilde{\theta_{0}}}(\boldsymbol{x}).$$

\State \textbf{return} $\left( \hat{{}\widetilde{{}\tau}}(\boldsymbol{x}), {\hat{{}\widetilde{\sigma_{1}}}}^{2}(\boldsymbol{x})\right)$
\end{algorithmic}
\end{algorithm}
\end{minipage}\\
%TC:endignore

Algorithm \ref{alg:EGRF} provides us with an estimate for both the CATE and the conditional variance of the ITE given the measured features. The ATE estimate remains the same as for the original CRF. The SD of the effect in the total population equals $$\sqrt{\mathbb{E}\left[\text{var}(Y^{1}-Y^{0} \mid \boldsymbol{X}) + \mathbb{E}\left[Y^{1}-Y^{0} \mid \boldsymbol{X} \right]^{2}\right] - \mathbb{E}[Y^{1}-Y^{0}]^{2}}$$ and is therefore estimated as 
$$\sqrt{\max\left\{0,n^{-1} \left(\sum_{i{=}1}^{n}  \hat{{}\widetilde{{}\sigma_{1}}}^{2}((\boldsymbol{x}_{i})) + \hat{{}\widetilde{{}\tau}}(\boldsymbol{x}_{i})^2 \right) - \widehat{\text{ATE}}^{2}\right\}}.$$

Only when the conditional ITE distribution can be well approximated with a Gaussian distribution the distribution of $Y^{1}-Y^{0}  \mid \boldsymbol{X}{=}\boldsymbol{x}$ is identified by the CATE and the conditional variance. Then, by the law of total probability, 
\begin{equation*}
\mathbb{P}(Y^1-Y^0 \leq y) ~{=}~ \int \mathbb{P}(Y^1-Y^0 \leq y  \mid \boldsymbol{X}{=}\boldsymbol{x}) dF_{\boldsymbol{X}}(\boldsymbol{x}).
\end{equation*} For illustration, we will assume the Gaussianity of the conditional ITE distribution in our example to use the extended CRF to estimate the ITE distribution from the simulated datasets.  The PEP is now estimated as $n^{-1}\sum_{i{=}1}^{n} \mathbb{P}(Z_{i}>0)$, where $Z_{i} \sim \mathcal{N}\left(\hat{{}\widetilde{{}\tau}}(\boldsymbol{x}_{i}), \max\left\{0,{\hat{{}\widetilde{\sigma_{1}}}}^{2}(\boldsymbol{x})\right\}\right)$.  

The Gaussianity assumption plays a different role than the identification assumptions \ref{CH2A1} to \ref{CH2A4}. The focus of this work is on the conditional variance (and the CATE) that is only identifiable under assumptions \ref{CH2A1} to \ref{CH2A4}. We resort to the Gaussianity assumption to also estimate the conditional effect distribution. In Section \ref{CH2CH2:sec5}, we will discuss that under violation of the Gaussianity assumption, the SD of the effect can still be appropriately estimated with Algorithm \ref{alg:EGRF}, but the PEP and ITE distribution estimates will be off. 

\subsection{Results resumed}
The bias, MSE and coverage for the ATE $(=0.5)$, SD $(=1.41)$ and PEP $(=0.64)$ of the ITE distribution, respectively, using the extended CRF while assuming Gaussian distributed $Y^{1}-Y^{0} \mid \boldsymbol{X}{=}\boldsymbol{x}$ are presented in Table \ref{CH2tab:2} for the different settings of the simulation study described in Section \ref{CH2sec:2.2}.

\begin{table}[H]
	\captionsetup{width=0.9\textwidth}
	\caption{Bias, mean squared error (MSE) and coverage of the estimated mean, SD and PEP ($\mathbb{P}(Y^1-Y^0>0)$) of the ITE distribution using the extended CRF based on $1000$ simulated samples per setting. 
 }\label{CH2tab:2}
	\centering
 \resizebox{0.85\textwidth}{!}{
	\begin{tabular}{cc||ccc|ccc|ccc}
		\multicolumn{2}{c||}{} & \multicolumn{3}{c|}{Bias} & \multicolumn{3}{c|}{MSE} & \multicolumn{3}{c}{Coverage} \\ \hline
		$\rho$    & $n$      & ATE    & SD     & PEP    & ATE    & SD     & PEP   & ATE      & SD      & PEP     \\ \hline
	0 & 200 & 0.05 & -0.06 & 0.04 & 0.20 & 0.25 & 0.02 & 0.95 & 0.90 & 0.91 \\ 
  0 & 2000 & -0.00 & 0.04 & 0.01 & 0.02 & 0.02 & 0.00 & 0.94 & 0.93 & 0.90 \\ 
  0 & 20000 & 0.00 & 0.03 & 0.01 & 0.00 & 0.00 & 0.00 & 0.95 & 0.89 & 0.82 \\ 
  0.25 & 200 & 0.01 & -0.08 & 0.03 & 0.20 & 0.29 & 0.02 & 0.94 & 0.88 & 0.92 \\ 
  0.25 & 2000 & 0.00 & 0.04 & 0.01 & 0.02 & 0.02 & 0.00 & 0.94 & 0.93 & 0.90 \\ 
  0.25 & 20000 & -0.00 & 0.04 & 0.00 & 0.00 & 0.00 & 0.00 & 0.95 & 0.86 & 0.86 \\ 
  0.50 & 200 & 0.02 & -0.07 & 0.03 & 0.19 & 0.27 & 0.02 & 0.95 & 0.90 & 0.92 \\ 
  0.50 & 2000 & 0.01 & 0.05 & 0.01 & 0.02 & 0.02 & 0.00 & 0.95 & 0.94 & 0.91 \\ 
  0.50 & 20000 & 0.00 & 0.05 & 0.00 & 0.00 & 0.00 & 0.00 & 0.94 & 0.80 & 0.92 \\ 
  0.75 & 200 & 0.04 & -0.07 & 0.03 & 0.16 & 0.26 & 0.02 & 0.95 & 0.92 & 0.92 \\ 
  0.75 & 2000 & 0.01 & 0.07 & -0.00 & 0.01 & 0.03 & 0.00 & 0.95 & 0.91 & 0.94 \\ 
  0.75 & 20000 & 0.00 & 0.07 & -0.00 & 0.00 & 0.01 & 0.00 & 0.94 & 0.71 & 0.93 \\ 
  1.00 & 200 & 0.04 & -0.03 & 0.02 & 0.14 & 0.26 & 0.02 & 0.95 & 0.91 & 0.92 \\ 
  1.00 & 2000 & 0.01 & 0.09 & -0.01 & 0.01 & 0.03 & 0.00 & 0.96 & 0.90 & 0.93 \\ 
  1.00 & 20000 & 0.00 & 0.09 & -0.01 & 0.00 & 0.01 & 0.00 & 0.95 & 0.56 & 0.89 \\ %\hline
	\end{tabular}%
 }
\end{table}
\noindent In the case of remaining heterogeneity ($\rho<1$), the bias of the SD estimator using the extended CRF is much lower than using the CRF. For larger sample sizes ($n=2000$ and $n=20000$), the small bias is of opposite sign to the one using the CRF. The bias of the PEP is also seriously decreased. For all settings, the MSE of the extended estimator is smaller for both the SD and PEP. Also, the coverage of the SD and PEP did considerably improve. However, for $n=20000$ and $\rho~{=}~0.50$ or $\rho~{=}~0.75$, the coverage probability of the SD did deviate from the nominal level due to the small bias and the narrow CIs. 

The extended CRF still performs well for the $\rho~{=}~1$ case, where all variability in causal effect could be explained with the measured features. Only when $n=20000$ the bias of the SD estimator using the extended CRF is slightly higher than for the original estimator due to an overspecified model. The difference is so small that the MSE is of the same magnitude. In this case, the coverage again deviates from the nominal level and is now slightly lower than the coverage using the traditional CRF. 

\begin{figure}[H]
	\centering
	\resizebox{0.9\linewidth}{!}{\includegraphics{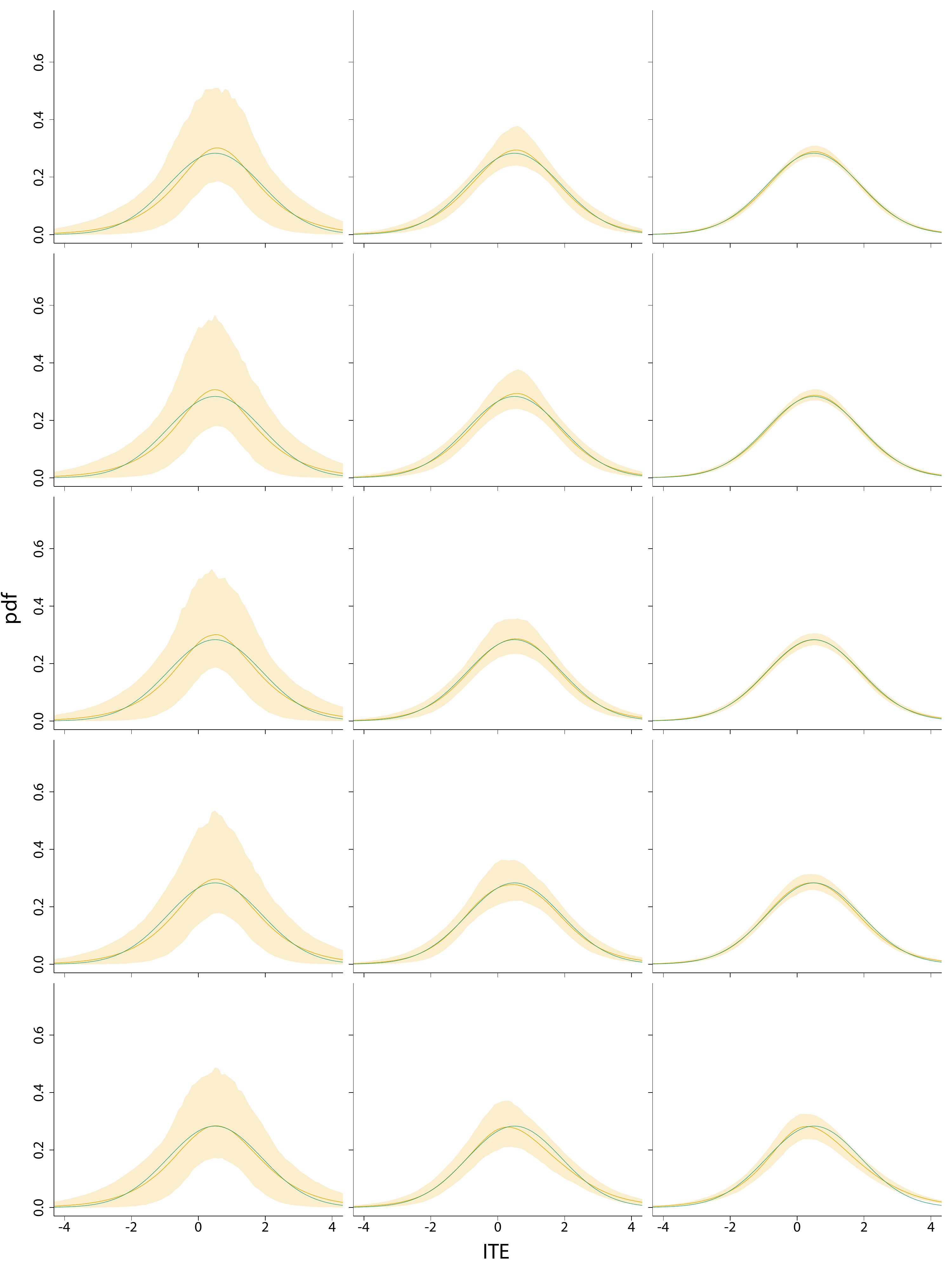}}
	\captionsetup{width=0.9\textwidth}
	\caption{Pointwise empirical mean (opaque yellow) as well as the $2.5\%$ and $97.5\%$ quantiles (transparent yellow) of the estimated ITE densities using the extended CRF of $1000$ simulations. Sample size per simulation varies over the columns and equals $200$ (left), $2000$ (middle) and $20000$ (right). The association of the measured $X_{0}$ and unmeasured modifier $U_{1}$ varies per row equal to $0, 0.25, 0.50, 0.75$ and $1$ (top to bottom). Furthermore, the actual ITE distribution is presented (green). }\label{CH2fig3}
\end{figure}

The pointwise mean (and $95\%$ CI), from $1000$ simulations, of the estimated probability density function of the ITE, is presented in Figure \ref{CH2fig3} %in Appendix \ref{CH2app:figd} 
for the different settings. %The extended CRF is expected to recover the ITE distribution well for all settings. 

\section{Concluding remarks}\label{CH2CH2:sec5}
ML methods are of great value in understanding effect heterogeneity using CATEs. In this work, we have shown that there might be individual effect modifications that cannot be explained by the features collected. As a result, the individualized CATE can still seriously differ from the ITE. Then, the ITE distribution cannot be identified by the distribution of the (random) conditional expectation alone, and applied researchers must be aware of this possible discrepancy. For example, remaining effect heterogeneity beyond heterogeneity in the CATEs can result in a lack of generalizability since the distribution of unmeasured effect modifiers in other populations might seriously differ from that in the sample \citep{Seamans2021}.  

Studying the remaining effect heterogeneity is challenging as the fundamental problem of causal inference prevents us from learning the joint distribution of potential outcomes. Nevertheless, the conditional second moments of the treated and the controls should be similar under remaining effect homogeneity. As an example, we have extended the CRF algorithm \citep{Athey2019} also to estimate the difference in conditional variance between treated and controls. If variances are different,  under assumptions \ref{CH2A1}, \ref{CH2A2}, and \ref{CH2A3}, the ITE distribution cannot be explained by the CATEs alone. The increased variance among the treated is due to the ITE's conditional variance and the covariance of the ITE and $Y^{0}$. Therefore, to estimate the (conditional) variance of the ITE, we need to assume how the ITE and $Y^{0}$ are correlated. In the examples presented in this work, $Y^{1}-Y^{0} \independent Y^{0}  \mid \boldsymbol{X}{=}\boldsymbol{x}$ so that the causal assumption of conditional independent effect deviation  applies. Under this assumption, the conditional variance of the ITE can be estimated next to the expected effect for each individual. As a result, in contrast to the CRF, the extended CRF can be used to estimate the ITE distribution's variance unbiasedly. It should be clear that for settings where Assumption \ref{CH2A4} is violated, the estimate of the (conditional) ITE variance based on the extended CRF will be biased as we illustrate with several scenarios as presented in Appendix \ref{app:B1}. The (conditional) variance of the ITE is also identifiable when the  conditional independent effect deviation is violated, but the joint distribution of $Y^1-Y^0$ and $Y^0$ is known. To identify the (conditional) variance of the ITE, the dependence structure of $Y^1-Y^0$ and $Y^0$ should be known since this cannot be learned from the data. The independence Assumption \ref{CH2A4} presented in this work is an example of this, but other assumptions on the dependence would also suffice. 

 In the absence of remaining effect heterogeneity, the estimated individual variance will be small, indicating that the CATE can be used as an appropriate proxy for the ITE. When assuming that the conditional ITE distributions can be approximated with Gaussian distributions, as done in our example, the ITE distribution can also be estimated. Note that when the conditional ITE distributions are not Gaussian, the distribution is not captured by the CATE and conditional variance alone. Then, other distributional properties like the PEP estimate will be off, as we have demonstrated for a scenario with non-Gaussian conditional treatment effects in Appendix \ref{app:B2}. However, the (conditional) ITE variance can still be appropriately estimated in this scenario. In this paper, we have focussed on the identification of the (conditional) variance of the causal effect. %It was only after assuming the Gaussianity of the conditional effect distribution that we could estimate other properties of the conditional distribution. 
However, under Assumption \ref{CH2A4} (and assumptions \ref{CH2A1}, \ref{CH2A2} and \ref{CH2A3}), also higher moments of the (conditional) ITE distribution are identifiable. One could derive the Robinson decomposition for $Y^3$ similarly to the derivation in Appendix \ref{CH2app:Y2RC}. 

 The effect modification relations in the example presented in the main paper were linear. However, as illustrated with the scenario presented in Appendix \ref{app:B3}, the linearity is not needed to estimate the (conditional) ITE variance using the extended CRF. We have presented the extended CRF just as an example, and other ML algorithms can be extended in a similar way to estimate the (conditional) variance of causal effects under Assumptions \ref{CH2A1}, \ref{CH2A2}, \ref{CH2A3} and \ref{CH2A4}.  This is only possible when the ML method appropriately estimates the CATEs and is not prone to overfitting \citep{Naimi2022, Balzer2022}. Furthermore, it is important to realize that the objective function considered is chosen to maximize heterogeneity in CATEs, and an algorithm might thus not account for confounders that are no effect modifiers (on the additive scale). Therefore, one should always discuss whether the distribution of potential outcomes is correctly linked to the observed distribution. In Appendix \ref{app:B4}, we present scenarios with confounders that are no effect modifiers (on the additive scale) to illustrate that for the CRF, the orthogonalization step facilitates this link. Without this step, even the ATE estimate can be biased. For studies 
where confounding is absent (e.g. a randomized experiment), the orthogonalization would thus not be necessary, as illustrated in the scenario presented in Appendix \ref{app:B5}

Although the extended CRF algorithm can be used in practice, our main aim was to emphasize that the CATE and ITE distributions can differ and present what assumption is necessary to identify the (conditional) variance of the causal effect. Since assumptions on the conditional dependence of $Y^0$ and the ITE cannot be tested with factual data, it will be challenging for epidemiologists and other experts in the field of an application of interest to review such assumptions. Judgement should be made based on reasoning about the causal pathways involved. It will be impossible for some applications to reason this way, but in others, it might be possible, like in the Clopidogrel example we mentioned in this paper. Reasoning about more examples will be an important topic of future interdisciplinary research. With this paper, we hope to open up the field of (conditional) ITE distribution estimation under assumptions like conditional independent effect deviation. The latter is necessary to fully understand to what degree individualized CATEs are informative at the actual individual level.  
	\bibliographystyle{spbasic}
	\bibliography{CEDistribution}

%TC:ignore

\newpage
\appendix

\captionsetup[table]{font=small}
\captionsetup[figure]{font=small}

\section{Robinson decomposition for squared outcome}\label{CH2app:Y2RC}
%\begin{equation}\mathbb{E}[\left(\tau(\boldsymbol{x}) + U_{1}\right)^{2}  \mid \boldsymbol{X}{=}\boldsymbol{x}] - \mathbb{E}[\tau(\boldsymbol{x})+ U_{1}  \mid \boldsymbol{X}{=}\boldsymbol{x}]^{2} ~{=}~\mathbb{E}[(U_{1})^{2}  \mid \boldsymbol{X}{=}\boldsymbol{x}].\end{equation}

First, we use the parameterization in \eqref{CH2eq:Y} to describe the conditional squared outcome, 
$$
(Y_{i})^2 = \left(\theta_{0}(\boldsymbol{X}_{i}) + N_{Yi} +  \left(\tau(\boldsymbol{X}_{i})+U_{1i} \right)A_{i} \right)^2.$$
For $\boldsymbol{X}_{i}=\boldsymbol{x}$, the conditional squared outcome is equal to 
$$\left(\theta_{0}(\boldsymbol{x}) + N_{Yi}\right)^2 +\left((\tau(\boldsymbol{x})+U_{1i})^{2} + 2(\tau(\boldsymbol{x})+U_{1i})(\theta_{0}(\boldsymbol{x})+N_{Yi}) \right)A_{i},$$
where the distributions of $A_{i}, N_{Yi}$ and $U_{1i}$ can depend on the value $\boldsymbol{x}$. The expression can be expanded as 
$$\left(\theta_{0}(\boldsymbol{x})\right)^{2} + N_{Yi}^2 + 2\theta_{0}(\boldsymbol{x})N_{Yi} + \left(\tau(\boldsymbol{x})^{2} + U_{1i}^{2} + 2\tau(\boldsymbol{x})U_{1i} + 2\tau(\boldsymbol{x})\theta_{0}(\boldsymbol{x}) + 2\tau(\boldsymbol{x})N_{Yi} + 2\theta_{0}(\boldsymbol{x})U_{1i} + 2N_{Yi}U_{1i} \right)A_{i}.$$ So that the conditional expectation of the squared outcome, $h(\boldsymbol{x}) = \mathbb{E}[(Y)^2  \mid  \boldsymbol{X}{=}\boldsymbol{x}]$, is equal to
%$$\left(\theta_{0}(\boldsymbol{x})\right)^{2} + \mathbb{E}[N_{Yi}^2 \mid \boldsymbol{X}_{i}=\boldsymbol{x}] + 2\theta_{0}(\boldsymbol{x})\mathbb{E}[N_{Yi} \mid \boldsymbol{X}_{i}=\boldsymbol{x}] + \left(\tau(\boldsymbol{x})^{2} + \mathbb{E}[U_{1i}^{2} \mid \boldsymbol{X}_{i}=\boldsymbol{x}] + 2\tau(\boldsymbol{x})\mathbb{E}[U_{1i} \mid \boldsymbol{X}_{i}=\boldsymbol{x}]+ \tau(\boldsymbol{x})\theta_{0}(\boldsymbol{x}) + \tau(\boldsymbol{x}) \mathbb{E}[N_{Yi}\mid \boldsymbol{X}_{i}=\boldsymbol{x}] + \theta_{0}(\boldsymbol{x})\mathbb{E}[U_{1i}\boldsymbol{X}_{i}=\boldsymbol{x}] + \mathbb{E}[N_{Yi}U_{1i} \mid \boldsymbol{X}_{i}=\boldsymbol{x}]\right) \mathbb{E}[A_{i}\boldsymbol{X}_{i}=\boldsymbol{x}].$$
\begin{align*}&\left(\theta_{0}(\boldsymbol{x})\right)^{2} + \mathbb{E}[N_{Yi}^2 \mid \boldsymbol{X}_{i}=\boldsymbol{x}] + \left(\tau(\boldsymbol{x})^{2} +  2\tau(\boldsymbol{x})\theta_{0}(\boldsymbol{x})\right) \mathbb{E}[A_{i}\mid \boldsymbol{X}_{i}=\boldsymbol{x}] +  \mathbb{E}[U_{1i}^{2}A_{i} \mid \boldsymbol{X}_{i}=\boldsymbol{x}] \\
& + 2(\tau(\boldsymbol{x})+ \theta_{0}(\boldsymbol{x}))\mathbb{E}[U_{1i}A_{i} \mid \boldsymbol{X}_{i}=\boldsymbol{x}] + 2\tau(\boldsymbol{x}) \mathbb{E}[N_{Yi}A_{i}\mid \boldsymbol{X}_{i}=\boldsymbol{x}] + 2\mathbb{E}[N_{Yi}U_{1i}A_{i} \mid \boldsymbol{X}_{i}=\boldsymbol{x}],\end{align*} since 
$\mathbb{E}[N_{Yi}\mid \boldsymbol{X}_{i}=\boldsymbol{x}]=0$. %as the mean of the potential outcome under no exposure given $\boldsymbol{X}_{i}=\boldsymbol{x}$ is described by $\theta_{0}(\boldsymbol{x})$. 
So, for $\boldsymbol{X}_{i}=\boldsymbol{x}$, $(Y_{i})^{2} - h(\boldsymbol{x})$, equals
\begin{align*} &(N_{Yi}^{2} - \mathbb{E}[N_{Yi}^{2} \mid \boldsymbol{X}_{i}=\boldsymbol{x}]) + 2 \theta_{0}(\boldsymbol{x})N_{Yi} + (A_{i}-\mathbb{E}[A_{i}\mid \boldsymbol{X}_{i}=\boldsymbol{x}])(\tau(\boldsymbol{x})^{2} + 2\tau(\boldsymbol{x})\theta_{0}(\boldsymbol{x}))\\
&+ (A_{i}U_{1i}^{2}- \mathbb{E}[A_{i}U_{1i}^{2}\mid \boldsymbol{X}_{i}=\boldsymbol{x}]) + 2(A_{i}U_{1i} - \mathbb{E}[A_{i}U_{1i}\mid \boldsymbol{X}_{i}=\boldsymbol{x} ])(\tau(\boldsymbol{x})+\theta_{0}(\boldsymbol{x}))\\ &+ 2\tau(\boldsymbol{x})(A_{i}N_{Yi}-\mathbb{E}[A_{i}N_{Yi}\mid \boldsymbol{X}_{i}=\boldsymbol{x} ]) + 2(A_{i}N_{Yi}U_{1i} - \mathbb{E}[A_{i}N_{Yi}U_{1i}\mid \boldsymbol{X}_{i}=\boldsymbol{x} ])
\end{align*}

In absence of unmeasured confounding, $A \independent N_{Y}, U_{1} \mid \boldsymbol{X}$, so that for $k$ equal to one or two $\mathbb{E}[(U_{1})^{k}A\mid \boldsymbol{X}{=}\boldsymbol{x}]=\mathbb{E}[(U_{1})^{k}]\mathbb{E}[A\mid\boldsymbol{X}{=}\boldsymbol{x}]$, $\mathbb{E}[A N_{Y}\mid\boldsymbol{X}{=}\boldsymbol{x}] = \mathbb{E}[A\mid\boldsymbol{X}{=}\boldsymbol{x}]\mathbb{E}[N_{Y}\mid\boldsymbol{X}{=}\boldsymbol{x}]$ and $\mathbb{E}[A N_{Y} U_{1}\mid\boldsymbol{X}{=}\boldsymbol{x}]=\mathbb{E}[A\mid\boldsymbol{X}{=}\boldsymbol{x}]\mathbb{E}[N_{Y} U_{1}\mid\boldsymbol{X}{=}\boldsymbol{x}]$. Then, by rewriting $A_{i} g(N_{Yi}, U_{1})$ as $A_{i} \mathbb{E}[g(N_{Yi} , U_{1})\mid\boldsymbol{X}_{i}=\boldsymbol{x}] + A_{i}(g(N_{Yi}, U_{1})-\mathbb{E}[g(N_{Yi}, U_{1})\mid\boldsymbol{X}_{i}=\boldsymbol{x}])$, for $\boldsymbol{X}_{i}=\boldsymbol{x}$, $$(Y_{i})^{2} - h(\boldsymbol{x})=\left(\tau(\boldsymbol{x})^{2} + %\mathbb{E}[U_{1i}^{2} \mid \boldsymbol{X}_{i}{=}\boldsymbol{x}] 
\sigma_{1}^{2}(\boldsymbol{x}) + 2\tau(\boldsymbol{x})\theta_{0}(\boldsymbol{x})  + 2\mathbb{E}[N_{Yi}U_{1i} \mid \boldsymbol{X}_{i}{=}\boldsymbol{x}]  \right)(A_{i}-\mathbb{E}[A_{i}\mid \boldsymbol{X}_{i}=\boldsymbol{x}]) + \gamma_{i}, $$
where
\begin{align*}\gamma_{i} = &(N_{Yi}^{2} - \mathbb{E}[N_{Yi}^{2} \mid \boldsymbol{X}_{i}=\boldsymbol{x}]) + 2 \theta_{0}(\boldsymbol{x})N_{Yi} \\ 
&+A_{i}\left( U_{1i}^{2}-%\mathbb{E}[U_{1i}^{2}\mid \boldsymbol{X}_{i}=\boldsymbol{x}])
\sigma_{1}^{2}(\boldsymbol{x}) 
 + 2(\tau(\boldsymbol{x})+\theta_{0}(\boldsymbol{x}))U_{1i}%-\mathbb{E}[U_{1i}\mid \boldsymbol{X}_{i}=\boldsymbol{x}])
+2\tau(\boldsymbol{x})N_{Yi}
+ 2(N_{Yi}U_{1i}-\mathbb{E}[N_{Yi}U_{1i}\mid \boldsymbol{X}_{i}=\boldsymbol{x}])\right),
\end{align*}
and $\forall \boldsymbol{x}{:}~ \mathbb{E}[\gamma_{i} \mid A{=}1, \boldsymbol{X}{=}\boldsymbol{x}]=\mathbb{E}[\gamma_{i} \mid A{=}0, \boldsymbol{X}{=}\boldsymbol{x}]=0$ as a result of the absence of unmeasured confounding. 

The Robinson decomposition estimand $\Delta(\boldsymbol{x})$ results from regressing the normalized squared outcome and the normalized treatment assignment, i.e.~ by fitting
\begin{equation*}
(Y_{i})^{2} - h(\boldsymbol{X}_{i}) = \left(A_{i} -e(\boldsymbol{x})\right) \Delta(\boldsymbol{x}) + \widetilde{{}\gamma}_{i},\end{equation*}
where $\forall \boldsymbol{x}{:}~ \mathbb{E}[\widetilde{{}\gamma}_{i} \mid A_{i}=1, \boldsymbol{X}_{i}=\boldsymbol{x}]=\mathbb{E}[\widetilde{{}\gamma}_{i} \mid A_{i}=0, \boldsymbol{X}_{i}=\boldsymbol{x}]=0$, 

In the absence of unmeasured confounding, 
$$\Delta(\boldsymbol{x}) = \tau(\boldsymbol{x})^{2} + %\mathbb{E}[U_{1i}^{2} \mid \boldsymbol{X}_{i}{=}\boldsymbol{x}]  
\sigma_{1}^{2}(\boldsymbol{x}) + 2\tau(\boldsymbol{x})\theta_{0}(\boldsymbol{x})  + 2\mathbb{E}[N_{Y}U_{1} \mid \boldsymbol{X}{=}\boldsymbol{x}],$$ and may be used to estimate the conditional variance of the causal effect. 

\newpage
\section{Other scenarios}\label{CH2app:Scenarios}
The cause-effect relations in Equation \eqref{eq:example}, presented in the main paper, were used as an example to illustrate the remaining heterogeneity beyond heterogeneity in CATEs. We have shown that under the assumption of conditional independent effect deviation, the variance of the conditional effect distribution can be identified. Moreover, when the conditional effect distribution is Gaussian, the estimates for the CATE and the conditional variance identify the entire distribution. In this Appendix, we present different scenarios where the assumptions are not met. Furthermore, we present an example where the effect modification by measured features is non-linear, an example with a confounder that is no effect modifier (on the additive scale) and an example without confounding.  

\subsection{Dependent $Y^{0}$ and $Y^{1}-Y^{0}$}\label{app:B1}
Cause-effect relations in Equation \eqref{eq:example}, can be generalized with
    $$\begin{pmatrix}
        N_{Y} \\
        U_{1}
    \end{pmatrix} \sim \mathcal{N}\left(   \begin{pmatrix}
        0 \\
        0
    \end{pmatrix}, \begin{pmatrix}
    \sigma_{0}^{2} & \kappa \sigma_{0}\sigma_{1} \\
    \kappa \sigma_{0}\sigma_{1} & \sigma_{1}^{2} 
    \end{pmatrix} \right),$$
 so that the original example is obtained for $\kappa=0$ and the conditional independent effect deviation assumption is violated otherwise. The conditional variance of $Y^{1}$ is equal to that of $N_{Y}+U_{1}$ and thus equals $\sigma_{0}^{2} + \sigma_{1}^{2} + 2 \kappa \sigma_{0}\sigma_{1}$. While varying $\kappa$, keeping $\sigma_{0}=1.6$, we restrict the parameter values of $\sigma_{1}$ so that the conditional variance of $Y^{1}$ remains equal to $1.6^{2}+1.4^{2}$ as in the main example since the latter was based on a real case study. Thus, 
$$ \sigma_{1} = -1.6\kappa + \sqrt{1.6^{2}\kappa^{2} + 1.4^{2}},$$
while the values for all other parameters in the cause-effect relations remain the same as in the main example. 

%ACE remains same
%SDs become 3.73 2.42 1.41 0.83 0.55
%P(>0)s become 0.55 0.58 0.64 0.73 0.82
\subsubsection{$\kappa=0$}
First, we again present the bias, MSE and coverage of the estimated mean, SD and PEP for the case where $\kappa=0$ so that $Y^{1}-Y^{0}\independent Y^{0}$ and thus the conditional independent effect deviation assumption is satisfied. For this setting, the true values of the mean, SD and PEP of the ITE were equal to $0.5$, $1.41$ and $0.64$, respectively. The performance of the estimators based on the original GRF (Table \ref{CH2tab:D0}) and the generalized GRF (Table \ref{CH2tab:D0E}) based on $1000$ simulations can be found below for $n=2000$ and $\rho \in \{0, 0.5, 1\}$. These results were already presented in Table \ref{CH2tab:1} and Table \ref{CH2tab:2}, but since we used $1000$ new simulations per setting, the results may vary slightly. 

\begin{figure}[H]
	\centering
	\resizebox{0.9\linewidth}{!}{\includegraphics{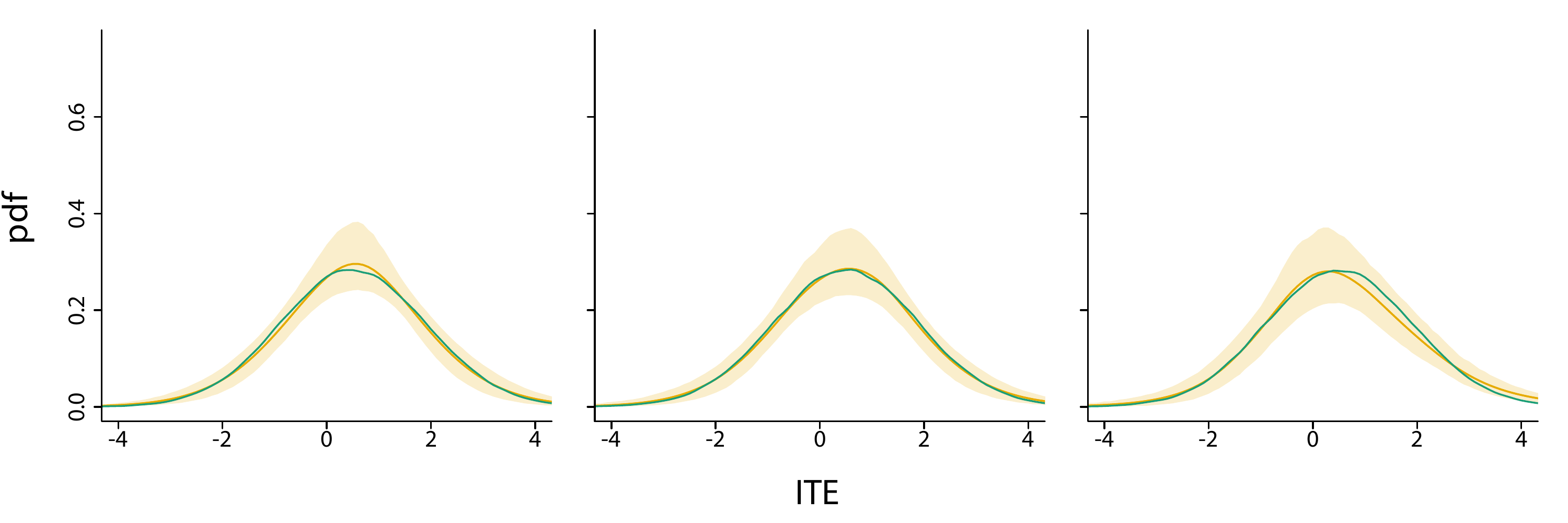}}
	\captionsetup{width=0.9\textwidth}
	\caption{Pointwise empirical mean (opaque yellow) as well as the $2.5\%$ and $97.5\%$ quantiles (transparent yellow) of the estimated ITE densities using the extended CRF of $1000$ simulations. The association of the measured $X_{0}$ and unmeasured modifier $U_{1}$ varies per column equal to $0, 0.50$ and $1$ (left to right), and the sample size per simulation equals $2000$. Furthermore, the actual ITE distribution $\kappa=0$ is presented (green). }\label{CH2D0fig3}
\end{figure}

\begin{table}[H]
	\captionsetup{width=0.9\textwidth}
	\caption{Bias, mean squared error (MSE) and coverage of the estimated mean, SD and PEP ($\mathbb{P}(Y^1-Y^0>0)$) of the ITE distribution using the characteristics of the CATE distribution as an estimator based on $1000$ simulated samples for $n=2000$, $\kappa=0$ and varying $\rho$. 
 }\label{CH2tab:D0}
	\centering
% \resizebox{0.85\textwidth}{!}{
	\begin{tabular}{cc||ccc|ccc|ccc}
		\multicolumn{2}{c||}{} & \multicolumn{3}{c|}{Bias} & \multicolumn{3}{c|}{MSE} & \multicolumn{3}{c}{Coverage} \\ \hline
		$\kappa$    & $\rho$      & ATE    & SD     & PEP    & ATE    & SD     & PEP   & ATE      & SD      & PEP     \\ \hline
0.00 & 0.00 & 0.00 & -1.18 & 0.34 & 0.02 & 1.39 & 0.12 & 0.95 & 0.00 & 0.21 \\ 
  0.00 & 0.50 & 0.01 & -0.82 & 0.16 & 0.02 & 0.69 & 0.03 & 0.96 & 0.00 & 0.56 \\ 
  0.00 & 1.00 & 0.00 & -0.21 & 0.01 & 0.01 & 0.06 & 0.00 & 0.97 & 0.41 & 0.94 \\ 
	\end{tabular}%
% }
\end{table} When studying the impact of dependence between $Y^{1}-Y^{0}$ and $Y^{0}$, it is important to remember that even for $\rho=1$, we observed a small finite sample bias for the SD estimator for $n=2000$. 
\begin{figure}[H]
	\centering
 	\captionsetup{width=0.9\textwidth}
	\resizebox{0.9\linewidth}{!}{\includegraphics{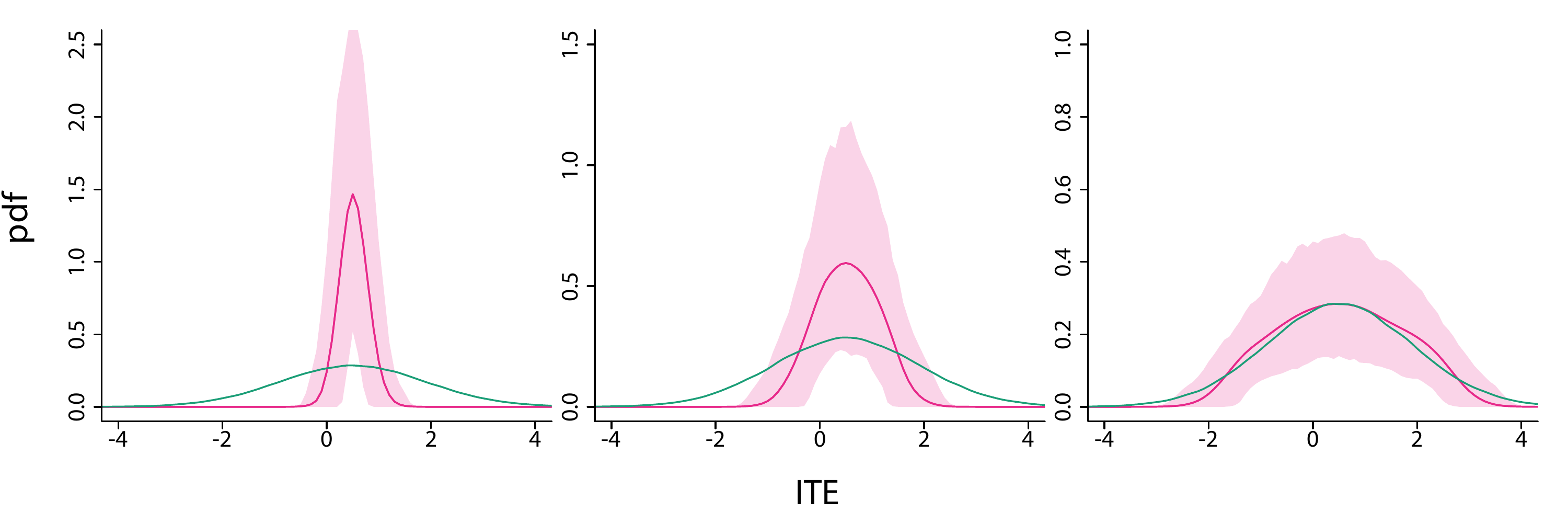}}
	\caption{Pointwise empirical mean (opaque pink) as well as the $2.5\%$ and $97.5\%$ quantiles (transparent pink) of the estimated ITE densities of $1000$ simulations. The association of the measured $X_{0}$ and unmeasured modifier $U_{1}$ varies per column equal to $0, 0.50$ and $1$ (left to right), and the sample size per simulation equals $2000$. Furthermore, the actual ITE distribution $\kappa=0$ is presented (green). }\label{CH2D0fig2}
\end{figure}
 The actual ITE distribution is presented together with the estimated ITE densities using the original CRF (Figure \ref{CH2D0fig2}) and the extended CRF (Figure \ref{CH2D0fig3}). Realize that in Figures \ref{CH2fig2} and \ref{CH2fig3}, different $\rho$ were represented by different rows instead of columns. 
\begin{table}[H]
	\captionsetup{width=0.9\textwidth}
	\caption{Bias, mean squared error (MSE) and coverage of the estimated mean, SD and PEP ($\mathbb{P}(Y^1-Y^0>0)$) of the ITE distribution using the extended CRF based on $1000$ simulated samples for $n=2000$, $\kappa=0$ and varying $\rho$. 
 }\label{CH2tab:D0E}
	\centering
% \resizebox{0.85\textwidth}{!}{
	\begin{tabular}{cc||ccc|ccc|ccc}
		\multicolumn{2}{c||}{} & \multicolumn{3}{c|}{Bias} & \multicolumn{3}{c|}{MSE} & \multicolumn{3}{c}{Coverage} \\ \hline
		$\kappa$    & $\rho$      & ATE    & SD     & PEP    & ATE    & SD     & PEP   & ATE      & SD      & PEP     \\ \hline
 0.00 & 0.00 & 0.00 & 0.04 & 0.01 & 0.02 & 0.02 & 0.00 & 0.95 & 0.92 & 0.91 \\ 
  0.00 & 0.50 & 0.01 & 0.05 & 0.01 & 0.02 & 0.02 & 0.00 & 0.96 & 0.92 & 0.93 \\ 
  0.00 & 1.00 & 0.00 & 0.08 & -0.02 & 0.01 & 0.03 & 0.00 & 0.97 & 0.92 & 0.92 \\ 
	\end{tabular}%
% }
\end{table} 

\clearpage
\subsubsection{$\kappa<0$}
For $\kappa=-0.5$ and $\kappa=-1$ the ATE remains $0.5$. However, as the variance of $Y^{1}$ and $Y^{0}$ are the same as in the $\kappa=0$ case, because of the negative correlation between $Y^{1}-Y^{0}$ and $Y^{0}$, the distribution of $Y^{1}-Y^{0}$ is higher in variability. The SD of the $Y^{1}-Y^{0}$ is now equal to $2.42$ ($\kappa=-0.5$) and $3.73$ ($\kappa=-1$) and the PEP equals $0.58$ ($\kappa=-0.5)$ and $0.55$ ($\kappa=-1$). 

\begin{table}[H]
	\captionsetup{width=0.9\textwidth}
	\caption{Bias, mean squared error (MSE) and coverage of the estimated mean, SD and PEP ($\mathbb{P}(Y^1-Y^0>0)$) of the ITE distribution using the characteristics of the CATE distribution as an estimator based on $1000$ simulated samples for $n=2000$, $\kappa\in\{-1, -0.5\}$ and $\rho\in\{0, 0.5, 1\}$. 
 }\label{CH2tab:D-1}
	\centering
 %\resizebox{0.85\textwidth}{!}{
	\begin{tabular}{cc||ccc|ccc|ccc}
		\multicolumn{2}{c||}{} & \multicolumn{3}{c|}{Bias} & \multicolumn{3}{c|}{MSE} & \multicolumn{3}{c}{Coverage} \\ \hline
		$\kappa$    & $\rho$      & ATE    & SD     & PEP    & ATE    & SD     & PEP   & ATE      & SD      & PEP     \\ \hline
 -0.50 & 0.00 & 0.00 & -2.18 & 0.39 & 0.02 & 4.77 & 0.15 & 0.95 & 0.00 & 0.10 \\ 
  -0.50 & 0.50 & 0.01 & -0.94 & 0.04 & 0.02 & 0.90 & 0.00 & 0.95 & 0.00 & 0.84 \\ 
  -0.50 & 1.00 & 0.01 & -0.26 & 0.01 & 0.01 & 0.08 & 0.00 & 0.95 & 0.10 & 0.94 \\ 
-1.00 & 0.00 & -0.00 & -3.49 & 0.42 & 0.02 & 12.21 & 0.18 & 0.96 & 0.00 & 0.06 \\ 
  -1.00 & 0.50 & 0.01 & -0.91 & 0.02 & 0.01 & 0.83 & 0.00 & 0.96 & 0.00 & 0.90 \\ 
  -1.00 & 1.00 & 0.01 & -0.34 & 0.01 & 0.01 & 0.12 & 0.00 & 0.96 & 0.00 & 0.88 \\  %\hline
	\end{tabular}%
% }
\end{table}

\begin{figure}[H]
	\centering
 	\captionsetup{width=0.9\textwidth}
	\resizebox{0.9\linewidth}{!}{\includegraphics{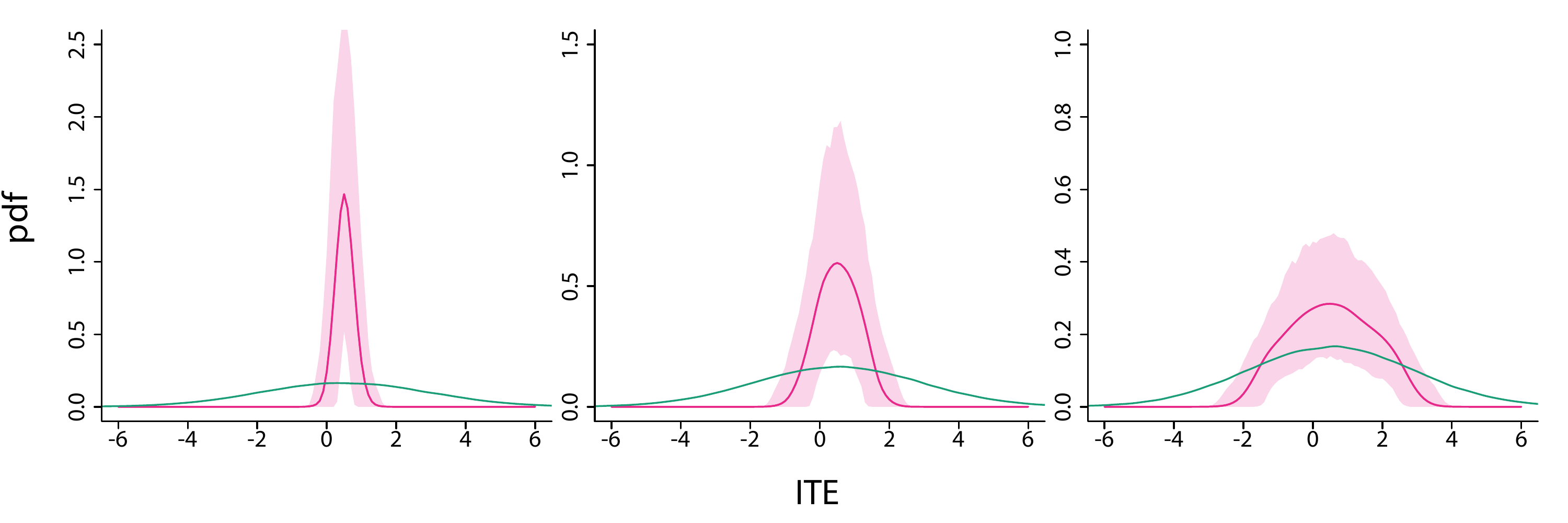}}
	\caption{Pointwise empirical mean (opaque pink) as well as the $2.5\%$ and $97.5\%$ quantiles (transparent pink) of the estimated ITE densities of $1000$ simulations. The association of the measured $X_{0}$ and unmeasured modifier $U_{1}$ varies per column equal to $0, 0.50$ and $1$ (left to right), and the sample size per simulation equals $2000$. Furthermore, the actual ITE distribution for $\kappa=-0.5$ is presented (green). }\label{CH2D-0.5fig2}
\end{figure}

\begin{figure}[H]
	\centering
 	\captionsetup{width=0.9\textwidth}
	\resizebox{0.9\linewidth}{!}{\includegraphics{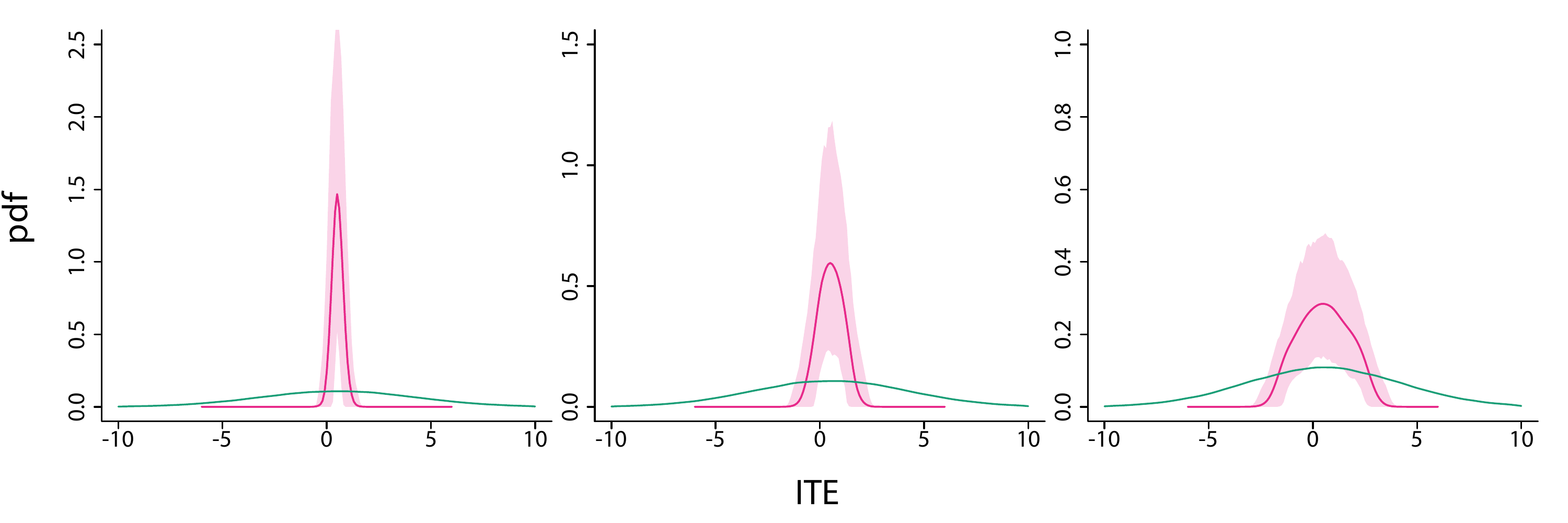}}
	\caption{Pointwise empirical mean (opaque pink) as well as the $2.5\%$ and $97.5\%$ quantiles (transparent pink) of the estimated ITE densities of $1000$ simulations. The association of the measured $X_{0}$ and unmeasured modifier $U_{1}$ varies per column equal to $0, 0.50$ and $1$ (left to right), and the sample size per simulation equals $2000$. Furthermore, the actual ITE distribution  for $\kappa=-1$ is presented (green). }\label{CH2D-1fig2}
\end{figure}

The estimand of the extended CRF equals $\widetilde{\sigma_{1}}^{2}(\boldsymbol{x})= \sigma_{1}^{2}(\boldsymbol{x}) + 2\mathbb{E}[N_{Y} U_{1}  \mid \boldsymbol{X}{=}\boldsymbol{x}]$. When the algorithm is used to estimate the (conditional) variance of the ITE distribution under Assumption \ref{CH2A4} while this assumption is violated because of the negative correlation, the variance is thus underestimated as presented in Table \ref{CH2tab:D-1E}. 
\begin{table}[H]
	\captionsetup{width=0.9\textwidth}
	\caption{Bias, mean squared error (MSE) and coverage of the estimated mean, SD and PEP ($\mathbb{P}(Y^1-Y^0>0)$) of the ITE distribution using the extended CRF based on $1000$ simulated samples for $n=2000$, $\kappa\in\{-1, -0.5\}$ and $\rho\in\{0, 0.5, 1\}$. 
 }\label{CH2tab:D-1E}
	\centering
 %\resizebox{0.85\textwidth}{!}{
	\begin{tabular}{cc||ccc|ccc|ccc}
		\multicolumn{2}{c||}{} & \multicolumn{3}{c|}{Bias} & \multicolumn{3}{c|}{MSE} & \multicolumn{3}{c}{Coverage} \\ \hline
		$\kappa$    & $\rho$      & ATE    & SD     & PEP    & ATE    & SD     & PEP   & ATE      & SD      & PEP     \\ \hline
 -0.50 & 0.00 & 0.00 & -0.96 & 0.07 & 0.02 & 0.94 & 0.01 & 0.95 & 0.00 & 0.49 \\ 
  -0.50 & 0.50 & 0.01 & -0.54 & 0.01 & 0.02 & 0.31 & 0.00 & 0.95 & 0.00 & 0.89 \\ 
  -0.50 & 1.00 & 0.01 & -0.16 & -0.01 & 0.01 & 0.04 & 0.00 & 0.95 & 0.56 & 0.94 \\
-1.00 & 0.00 & -0.00 & -2.27 & 0.10 & 0.02 & 5.19 & 0.01 & 0.96 & 0.00 & 0.22 \\ 
  -1.00 & 0.50 & 0.01 & -0.71 & 0.00 & 0.01 & 0.52 & 0.00 & 0.96 & 0.00 & 0.92 \\ 
  -1.00 & 1.00 & 0.01 & -0.23 & -0.00 & 0.01 & 0.06 & 0.00 & 0.96 & 0.02 & 0.95 \\ 
	\end{tabular}%
% }
\end{table}

\begin{figure}[H]
	\centering
	\resizebox{0.9\linewidth}{!}{\includegraphics{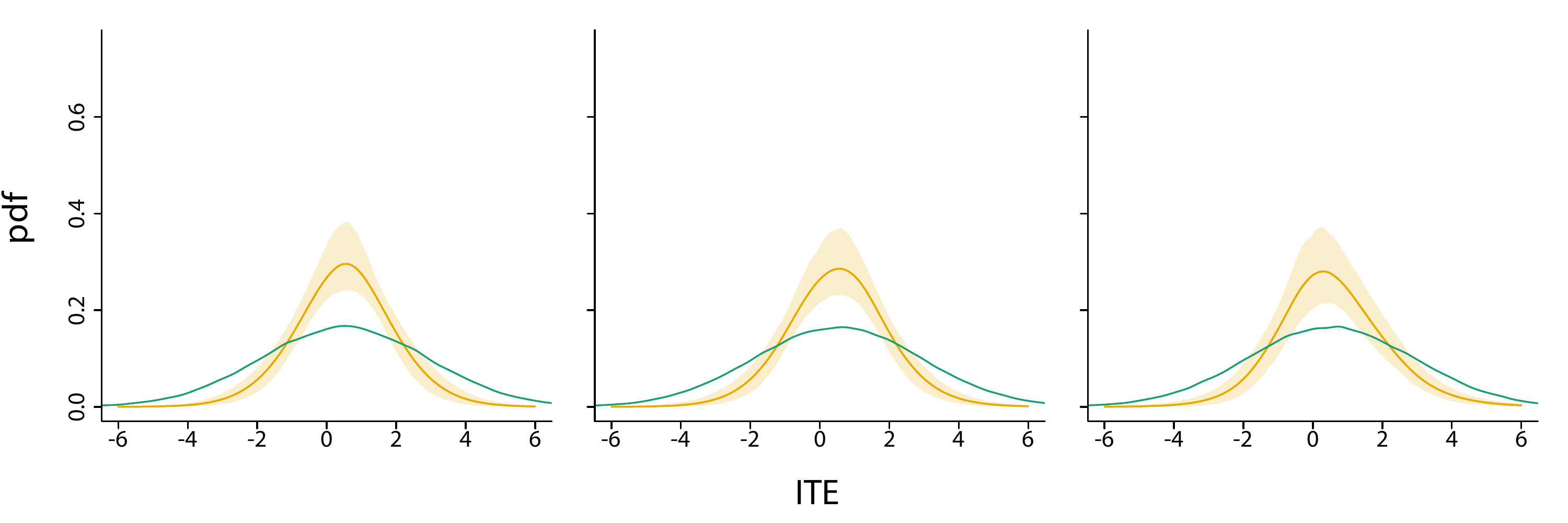}}
	\captionsetup{width=0.9\textwidth}
	\caption{Pointwise empirical mean (opaque yellow) as well as the $2.5\%$ and $97.5\%$ quantiles (transparent yellow) of the estimated ITE densities using the extended CRF of $1000$ simulations. The association of the measured $X_{0}$ and unmeasured modifier $U_{1}$ varies per column equal to $0, 0.50$ and $1$ (left to right), and the sample size per simulation equals $2000$. Furthermore, the actual ITE distribution $\kappa=-0.5$ is presented (green). }\label{CH2D-0.5fig3}
\end{figure}

\begin{figure}[H]
	\centering
	\resizebox{0.9\linewidth}{!}{\includegraphics{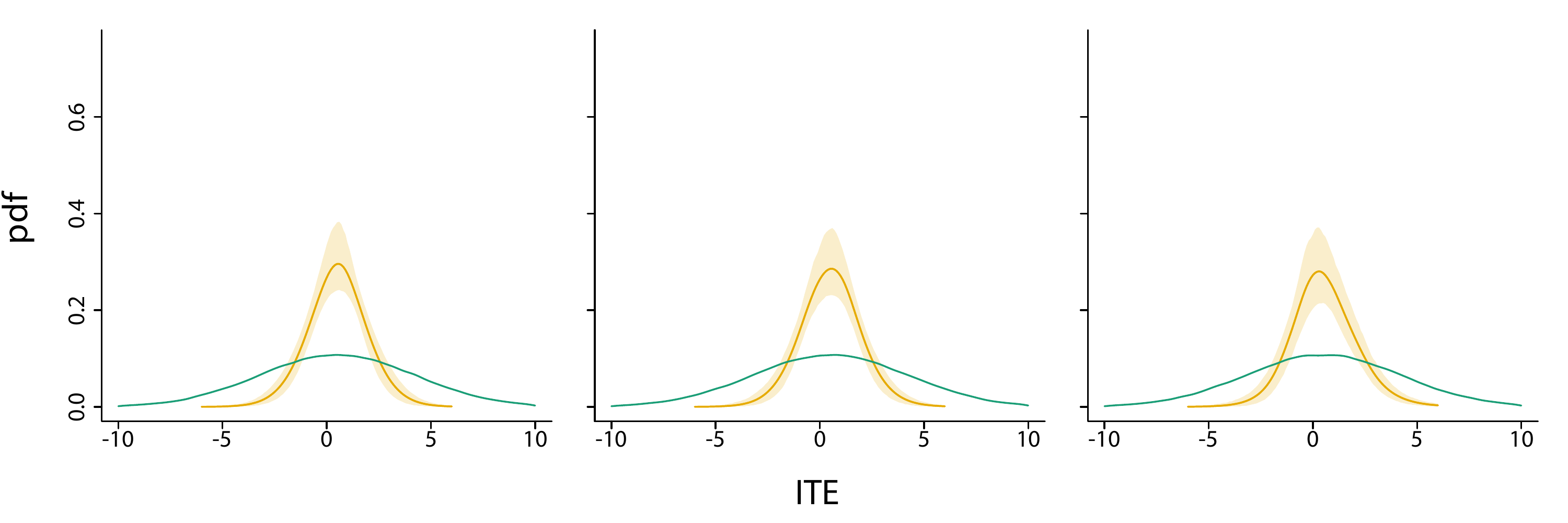}}
	\captionsetup{width=0.9\textwidth}
	\caption{Pointwise empirical mean (opaque yellow) as well as the $2.5\%$ and $97.5\%$ quantiles (transparent yellow) of the estimated ITE densities using the extended CRF of $1000$ simulations. The association of the measured $X_{0}$ and unmeasured modifier $U_{1}$ varies per column equal to $0, 0.50$ and $1$ (left to right), and the sample size per simulation equals $2000$. Furthermore, the actual ITE distribution $\kappa=-1$ is presented (green). }\label{CH2D-1fig3}
\end{figure}

\clearpage
\subsubsection{$\kappa>0$}
For $\kappa=0.5$ and $\kappa=1$ the ATE remains $0.5$. However, as the variance of $Y^{1}$ and $Y^{0}$ are the same as in the $\kappa=0$ case, because of the positive correlation between $Y^{1}-Y^{0}$ and $Y^{0}$, the distribution of $Y^{1}-Y^{0}$ is lower in variability. The SD of the $Y^{1}-Y^{0}$ is now equal to $0.83$ ($\kappa=0.5$) and $0.55$ ($\kappa=1$) and the PEP equals $0.73$ ($\kappa=0.5)$ and $0.82$ ($\kappa=1$). 

\begin{table}[H]
	\captionsetup{width=0.9\textwidth}
	\caption{Bias, mean squared error (MSE) and coverage of the estimated mean, SD and PEP ($\mathbb{P}(Y^1-Y^0>0)$) of the ITE distribution using the characteristics of the CATE distribution as an estimator based on $1000$ simulated samples for $n=2000$, $\kappa\in\{0.5, 1\}$ and $\rho\in\{0, 0.5, 1\}$. 
 }\label{CH2tab:D0.5}
	\centering
% \resizebox{0.85\textwidth}{!}{
	\begin{tabular}{cc||ccc|ccc|ccc}
		\multicolumn{2}{c||}{} & \multicolumn{3}{c|}{Bias} & \multicolumn{3}{c|}{MSE} & \multicolumn{3}{c}{Coverage} \\ \hline
		$\kappa$    & $\rho$      & ATE    & SD     & PEP    & ATE    & SD     & PEP   & ATE      & SD      & PEP     \\ \hline
  0.50 & 0.00 & 0.01 & -0.59 & 0.25 & 0.02 & 0.36 & 0.06 & 0.96 & 0.00 & 0.42 \\ 
  0.50 & 0.50 & 0.01 & -0.54 & 0.22 & 0.02 & 0.30 & 0.05 & 0.94 & 0.00 & 0.48 \\ 
  0.50 & 1.00 & 0.01 & -0.14 & 0.02 & 0.01 & 0.03 & 0.00 & 0.95 & 0.58 & 0.92 \\ 
    1.00 & 0.00 & 0.00 & -0.32 & 0.15 & 0.02 & 0.10 & 0.03 & 0.94 & 0.18 & 0.74 \\ 
  1.00 & 0.50 & 0.00 & -0.31 & 0.15 & 0.02 & 0.10 & 0.03 & 0.95 & 0.18 & 0.73 \\ 
  1.00 & 1.00 & 0.01 & -0.07 & 0.03 & 0.00 & 0.01 & 0.00 & 0.95 & 0.00 & 0.53 \\ 
	\end{tabular}%
% }
\end{table}
  \begin{figure}[H]
	\centering
 	\captionsetup{width=0.9\textwidth}
	\resizebox{0.9\linewidth}{!}{\includegraphics{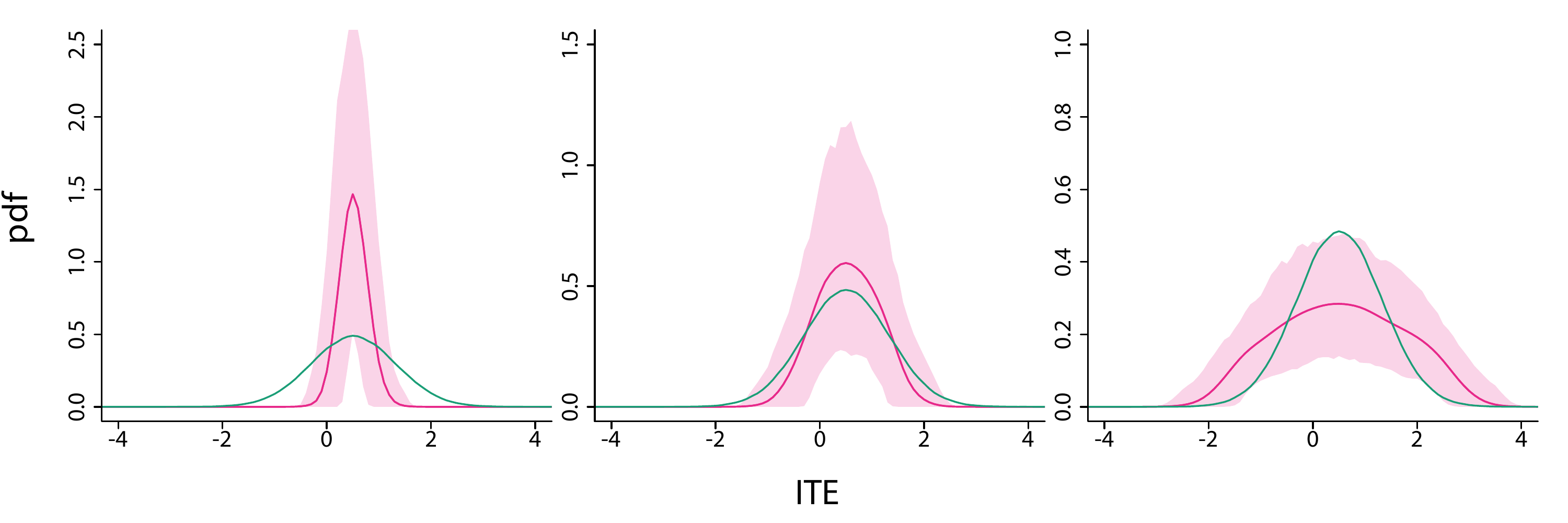}}
	\caption{Pointwise empirical mean (opaque pink) as well as the $2.5\%$ and $97.5\%$ quantiles (transparent pink) of the estimated ITE densities of $1000$ simulations. The association of the measured $X_{0}$ and unmeasured modifier $U_{1}$ varies per column equal to $0, 0.50$ and $1$ (left to right), and the sample size per simulation equals $2000$. Furthermore, the actual ITE distribution for $\kappa=0.5$ is presented (green). }\label{CH2D0.5fig2}
\end{figure}

\begin{figure}[H]
	\centering
 	\captionsetup{width=0.9\textwidth}
	\resizebox{0.9\linewidth}{!}{\includegraphics{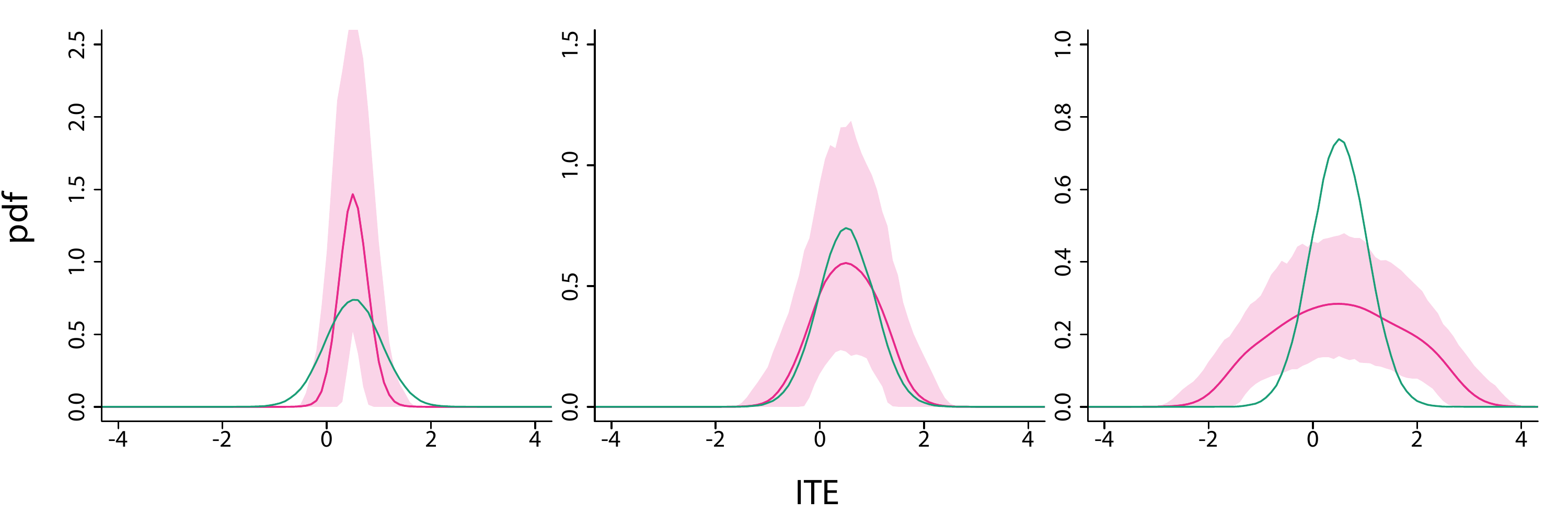}}
	\caption{Pointwise empirical mean (opaque pink) as well as the $2.5\%$ and $97.5\%$ quantiles (transparent pink) of the estimated ITE densities of $1000$ simulations. The association of the measured $X_{0}$ and unmeasured modifier $U_{1}$ varies per column equal to $0, 0.50$ and $1$ (left to right), and the sample size per simulation equals $2000$. Furthermore, the actual ITE distribution  for $\kappa=1$ is presented (green). }\label{CH2D1fig2}
\end{figure}
The estimand of the extended CRF equals $\widetilde{\sigma_{1}}^{2}(\boldsymbol{x})= \sigma_{1}^{2}(\boldsymbol{x}) + 2\mathbb{E}[N_{Y} U_{1}  \mid \boldsymbol{X}{=}\boldsymbol{x}]$. When the algorithm is used to estimate the (conditional) variance of the ITE distribution under Assumption \ref{CH2A4} while this assumption is violated because of the positive correlation, the variance is thus overestimated as presented in Table \ref{CH2tab:D0.5E}. 
\begin{table}[H]
	\captionsetup{width=0.9\textwidth}
	\caption{Bias, mean squared error (MSE) and coverage of the estimated mean, SD and PEP ($\mathbb{P}(Y^1-Y^0>0)$) of the ITE distribution using the extended CRF based on $1000$ simulated samples for $n=2000$, $\kappa\in\{0.5, 1\}$ and $\rho\in\{0, 0.5, 1\}$.  
 }\label{CH2tab:D0.5E}
	\centering
% \resizebox{0.85\textwidth}{!}{
	\begin{tabular}{cc||ccc|ccc|ccc}
		\multicolumn{2}{c||}{} & \multicolumn{3}{c|}{Bias} & \multicolumn{3}{c|}{MSE} & \multicolumn{3}{c}{Coverage} \\ \hline
		$\kappa$    & $\rho$      & ATE    & SD     & PEP    & ATE    & SD     & PEP   & ATE      & SD      & PEP     \\ \hline
 0.50 & 0.00 & 0.01 & 0.63 & -0.08 & 0.02 & 0.41 & 0.01 & 0.96 & 0.02 & 0.58 \\ 
  0.50 & 0.50 & 0.01 & 0.60 & -0.07 & 0.02 & 0.38 & 0.01 & 0.94 & 0.02 & 0.60 \\ 
  0.50 & 1.00 & 0.01 & 0.26 & -0.05 & 0.01 & 0.08 & 0.00 & 0.95 & 0.61 & 0.80 \\ 
  1.00 & 0.00 & 0.00 & 0.91 & -0.17 & 0.02 & 0.84 & 0.03 & 0.94 & 0.00 & 0.04 \\ 
  1.00 & 0.50 & 0.00 & 0.89 & -0.17 & 0.02 & 0.81 & 0.03 & 0.95 & 0.00 & 0.03 \\ 
  1.00 & 1.00 & 0.01 & 0.33 & -0.08 & 0.00 & 0.11 & 0.01 & 0.95 & 0.00 & 0.00 \\ 
	\end{tabular}%
% }
\end{table}

\begin{figure}[H]
	\centering
	\resizebox{0.9\linewidth}{!}{\includegraphics{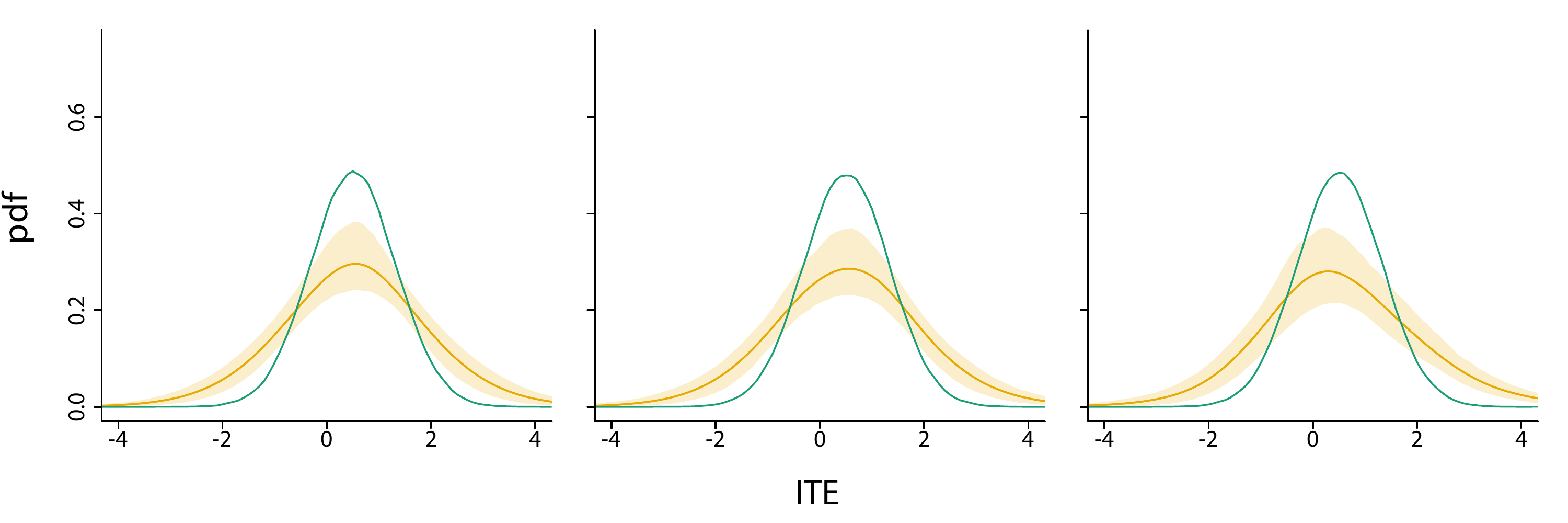}}
	\captionsetup{width=0.9\textwidth}
	\caption{Pointwise empirical mean (opaque yellow) as well as the $2.5\%$ and $97.5\%$ quantiles (transparent yellow) of the estimated ITE densities using the extended CRF of $1000$ simulations. The association of the measured $X_{0}$ and unmeasured modifier $U_{1}$ varies per column equal to $0, 0.50$ and $1$ (left to right), and the sample size per simulation equals $2000$. Furthermore, the actual ITE distribution $\kappa=0.5$ is presented (green). }\label{CH2D0.5fig3}
\end{figure}

\begin{figure}[H]
	\centering
	\resizebox{0.9\linewidth}{!}{\includegraphics{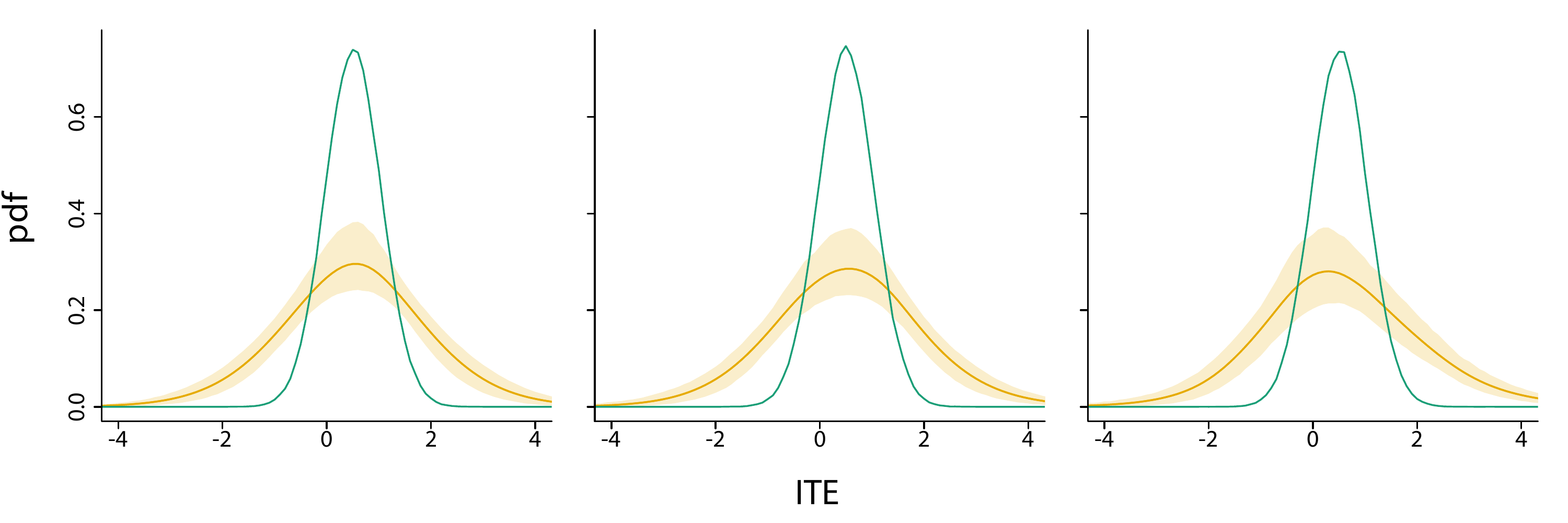}}
	\captionsetup{width=0.9\textwidth}
	\caption{Pointwise empirical mean (opaque yellow) as well as the $2.5\%$ and $97.5\%$ quantiles (transparent yellow) of the estimated ITE densities using the extended CRF of $1000$ simulations. The association of the measured $X_{0}$ and unmeasured modifier $U_{1}$ varies per column equal to $0, 0.50$ and $1$ (left to right), and the sample size per simulation equals $2000$. Furthermore, the actual ITE distribution $\kappa=1$ is presented (green). }\label{CH2D1fig3}
\end{figure}

\subsection{Non-Gaussian conditional effect distributions}\label{app:B2}
Also, a scenario with cause-effect relations in Equation \eqref{eq:example}, where $U_{1}$ follows a (shifted) log-normal distribution with again mean zero and variance $\sigma_{1}^{2}$ instead of Gaussian distributed $U_{1}$. More specifically, let
\begin{align}
    & \sigma = \sqrt{\log\left(0.5\sqrt{4\sigma_{1}^{2}+1}+0.5\right)}\\
    & \mu = \exp(0+0.5\sigma^{2}),
\end{align} and $U_{1} \overset{d}{=} \exp(X) - \mu$, where $X\sim \mathcal{N}(0,\sigma^{2})$. The values of all parameters are the same as in the main example. In this scenario, the joint distribution of $U_{1}$ and $X_{3}$ is no longer bivariate Gaussian as $U_{1}$ is not Gaussian distributed. Instead we assume that the conditional distribution of $X_{3i}$ given $U_{1i}=u_{1}$ is Gaussian distributed with mean $u_{1}\delta\rho $ and standard deviation $\delta \sigma_{1} \sqrt{1-\rho^{2}}$. The mean and SD of the ITE distribution remain $0.5$ and $1.41$, respectively, as in the main example. However, the PEP now equals $0.55$.

%ACE remains
%SD remains
% P0 becomes 0.55

\begin{table}[H]
	\captionsetup{width=0.9\textwidth}
	\caption{Bias, mean squared error (MSE) and coverage of the estimated mean, SD and PEP ($\mathbb{P}(Y^1-Y^0>0)$) of the ITE distribution using the characteristics of the CATE distribution as an estimator based on $1000$ simulated samples for $n=2000$, and $\rho \in \{0, 0.5, 1\}$. 
 }\label{CH2tab:LN}
	\centering
% \resizebox{0.85\textwidth}{!}{
	\begin{tabular}{c||ccc|ccc|ccc}
		 & \multicolumn{3}{c|}{Bias} & \multicolumn{3}{c|}{MSE} & \multicolumn{3}{c}{Coverage} \\ \hline
		 $\rho$      & ATE    & SD     & PEP    & ATE    & SD     & PEP   & ATE      & SD      & PEP     \\ \hline
0.00 & 0.00 & -1.17 & 0.42 & 0.02 & 1.38 & 0.18 & 0.95 & 0.00 & 0.06 \\ 
  0.50 & 0.01 & -0.87 & 0.33 & 0.02 & 0.77 & 0.12 & 0.95 & 0.00 & 0.23 \\ 
  1.00 & 0.01 & -0.34 & 0.06 & 0.01 & 0.13 & 0.01 & 0.95 & 0.22 & 0.88 \\ 
	\end{tabular}%
% }
\end{table} From the results in Table \ref{CH2tab:LN}, it becomes clear that the bias of the SD estimator is now similar to the bias in the case of Gaussian distributed conditional ITEs. For this setting, Assumption \ref{CH2A4} is valid so that the extended CRF can be used to estimate the SD as shown in Table \ref{CH2tab:LNE}. However, in the same table, it is illustrated that assuming Gaussianity of the conditional ITE, as in the primary example, now results in an overestimation of the PEP when $\rho$ equals $0$ or $0.5$. 
\begin{figure}[H]
	\centering
 	\captionsetup{width=0.9\textwidth}
	\resizebox{0.9\linewidth}{!}{\includegraphics{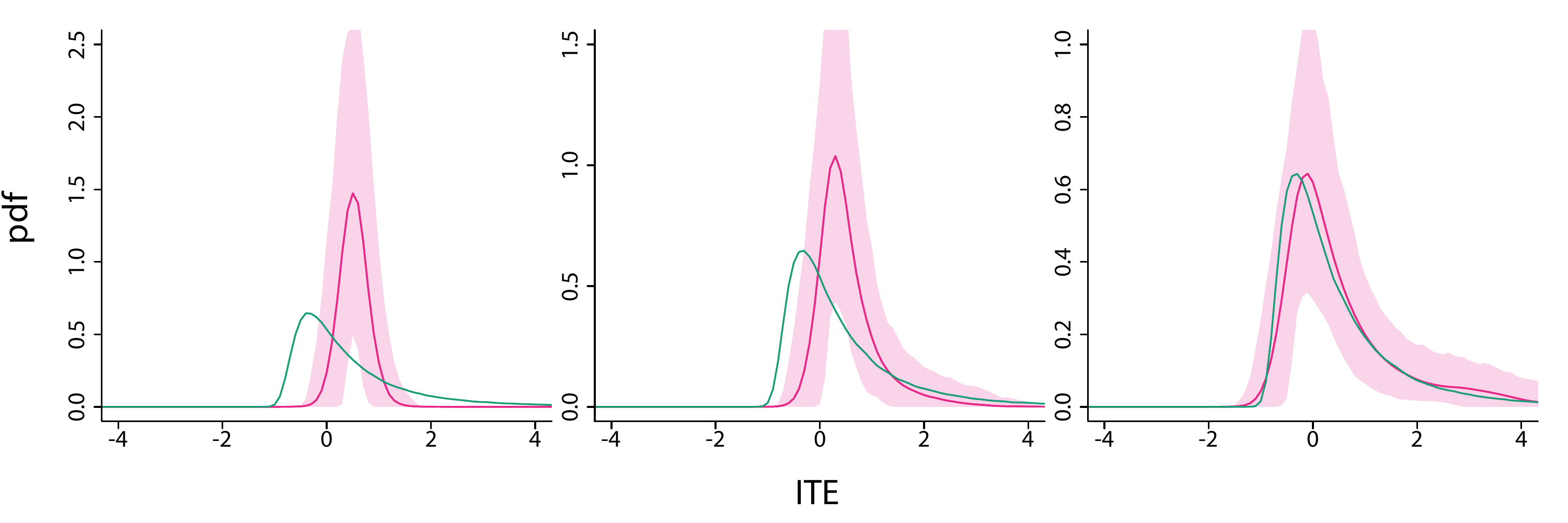}}
	\caption{Pointwise empirical mean (opaque pink) as well as the $2.5\%$ and $97.5\%$ quantiles (transparent pink) of the estimated ITE densities of $1000$ simulations. The association of the measured $X_{0}$ and unmeasured modifier $U_{1}$ varies per column equal to $0, 0.50$ and $1$ (left to right), and the sample size per simulation equals $2000$. Furthermore, the actual ITE distribution is presented (green). }\label{CH2LNfig2}
\end{figure}
\begin{table}[H]
	\captionsetup{width=0.9\textwidth}
	\caption{Bias, mean squared error (MSE) and coverage of the estimated mean, SD and PEP ($\mathbb{P}(Y^1-Y^0>0)$) of the ITE distribution using the extended CRF based on $1000$ simulated samples for $n=2000$ and $\rho \in \{0, 0.5, 1\}$. 
 }\label{CH2tab:LNE}
	\centering
% \resizebox{0.85\textwidth}{!}{
	\begin{tabular}{c||ccc|ccc|ccc}
		 & \multicolumn{3}{c|}{Bias} & \multicolumn{3}{c|}{MSE} & \multicolumn{3}{c}{Coverage} \\ \hline
		 $\rho$      & ATE    & SD     & PEP    & ATE    & SD     & PEP   & ATE      & SD      & PEP     \\ \hline
0.00 & 0.00 & 0.03 & 0.11 & 0.02 & 0.06 & 0.01 & 0.95 & 0.88 & 0.23 \\ 
  0.50 & 0.01 & 0.04 & 0.10 & 0.02 & 0.06 & 0.01 & 0.95 & 0.89 & 0.52 \\ 
  1.00 & 0.01 & 0.08 & 0.02 & 0.01 & 0.06 & 0.00 & 0.95 & 0.92 & 0.92 \\  
	\end{tabular}%
% }
\end{table} 
\begin{figure}[H]
	\centering
	\resizebox{0.9\linewidth}{!}{\includegraphics{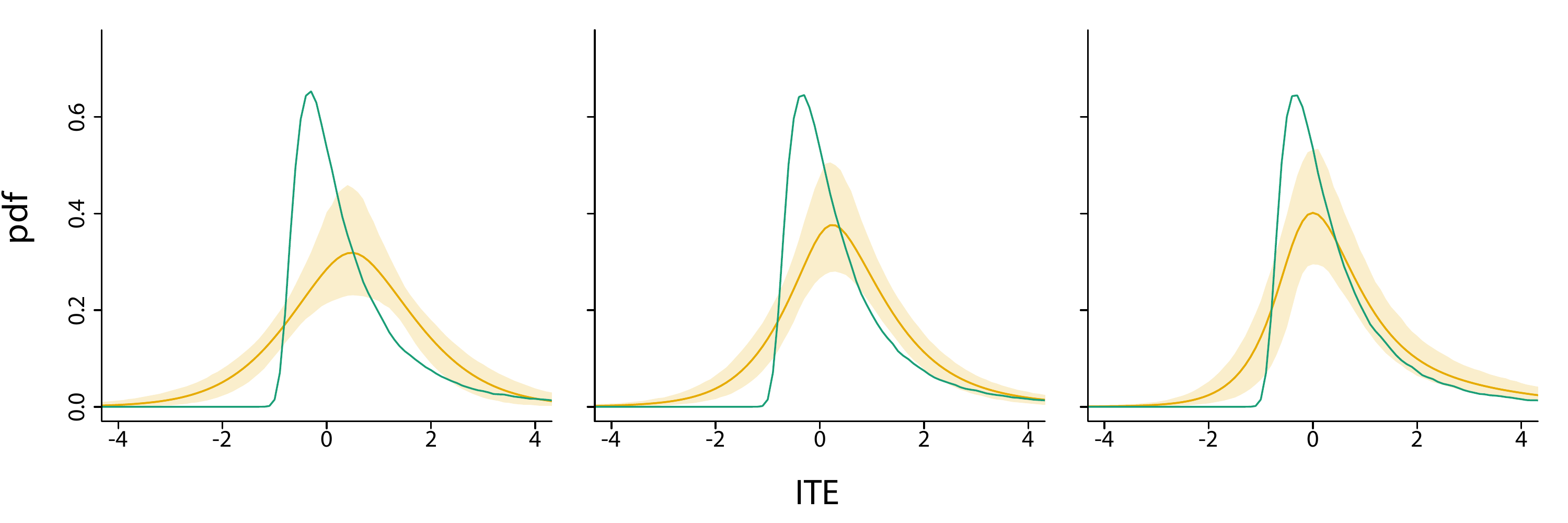}}
	\captionsetup{width=0.9\textwidth}
	\caption{Pointwise empirical mean (opaque yellow) as well as the $2.5\%$ and $97.5\%$ quantiles (transparent yellow) of the estimated ITE densities using the extended CRF of $1000$ simulations. The association of the measured $X_{0}$ and unmeasured modifier $U_{1}$ varies per column equal to $0, 0.50$ and $1$ (left to right), and the sample size per simulation equals $2000$. Furthermore, the actual ITE distribution is presented (green). }\label{CH2LNfig3}
\end{figure}

%\todo[inline]{Check Gaussianity with residuals Y1 <> Need a rather large N}

%\todo[inline]{Could we present an example of why the Gaussianity actually can be validated?}

\subsection{Non-linear effect modification}\label{app:B3}
As a scenario where the CATE is a nonlinear function of $X_{\text{sex}}$ and $X_{\text{SBP}}$ we adapt the cause-effect relations in Equation \eqref{eq:example} with
$$Y^{1}_{i}~{=}~ Y^{0}_{i} + \left( (\tau_{0} + \tau_{\text{sex}}X_{\text{sex},i})\exp(\tau_{\text{SBP}}X_{\text{SBP},i}) +U_{1i}\right).$$
New parameter values for $\tau_{\text{sex}}=0.10$ and $\tau_{\text{SBP}}=0.20$ where obtained by fitting 
$$ Y ~{=}~ \beta_{0} + \beta_{\text{sex}}X_{\text{sex},i} +\beta_{\text{SBP}}X_{\text{SBP},i} + N_{Yi} + 
 \left( (\tau_{0} + \tau_{\text{sex}}X_{\text{sex},i})\exp(\tau_{\text{SBP}}X_{\text{SBP},i}) +U_{1i}\right)A_{i}$$ to the subset of the FHS participants as used by \citet{Chiu2020} using \texttt{PROC NLMIXED} in \texttt{SAS}. All other parameters were taken the same as in the main example. The actual values for the mean, SD and PEP of the ITE distribution were also the same. 

 \begin{table}[H]
	\captionsetup{width=0.9\textwidth}
	\caption{Bias, mean squared error (MSE) and coverage of the estimated mean, SD and PEP ($\mathbb{P}(Y^1-Y^0>0)$) of the ITE distribution using the characteristics of the CATE distribution as an estimator based on $1000$ simulated samples for $n=2000$, and $\rho \in \{0, 0.5, 1\}$. 
 }\label{CH2tab:NL}
	\centering
% \resizebox{0.85\textwidth}{!}{
	\begin{tabular}{c||ccc|ccc|ccc}
		 & \multicolumn{3}{c|}{Bias} & \multicolumn{3}{c|}{MSE} & \multicolumn{3}{c}{Coverage} \\ \hline
		 $\rho$      & ATE    & SD     & PEP    & ATE    & SD     & PEP   & ATE      & SD      & PEP     \\ \hline
0.00 & 0.02 & -1.19 & 0.34 & 0.02 & 1.41 & 0.12 & 0.96 & 0.00 & 0.17 \\ 
  0.50 & 0.02 & -0.83 & 0.16 & 0.02 & 0.70 & 0.04 & 0.95 & 0.00 & 0.56 \\ 
  1.00 & 0.02 & -0.20 & 0.01 & 0.01 & 0.05 & 0.00 & 0.94 & 0.43 & 0.93 \\ 
	\end{tabular}%
% }
\end{table}

\begin{figure}[H]
	\centering
 	\captionsetup{width=0.9\textwidth}
	\resizebox{0.9\linewidth}{!}{\includegraphics{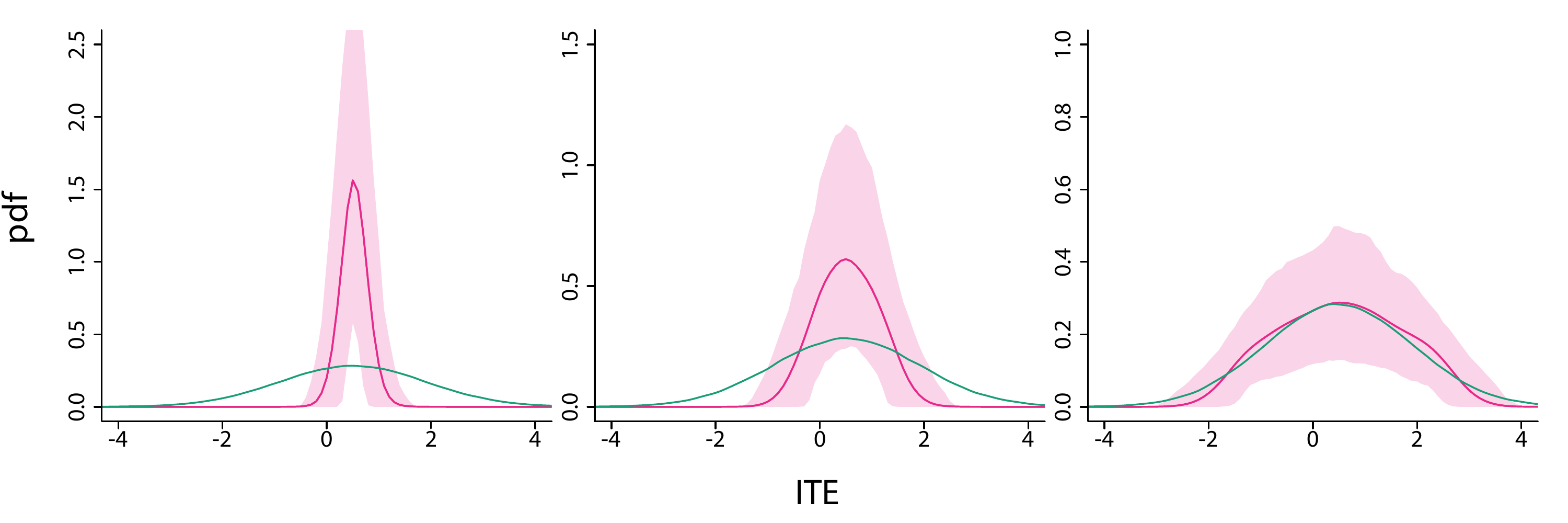}}
	\caption{Pointwise empirical mean (opaque pink) as well as the $2.5\%$ and $97.5\%$ quantiles (transparent pink) of the estimated ITE densities of $1000$ simulations. The association of the measured $X_{0}$ and unmeasured modifier $U_{1}$ varies per column equal to $0, 0.50$ and $1$ (left to right), and the sample size per simulation equals $2000$. Furthermore, the actual ITE distribution is presented (green). }\label{CH2NLfig2}
\end{figure}

\begin{table}[H]
	\captionsetup{width=0.9\textwidth}
	\caption{Bias, mean squared error (MSE) and coverage of the estimated mean, SD and PEP ($\mathbb{P}(Y^1-Y^0>0)$) of the ITE distribution using the extended CRF based on $1000$ simulated samples for $n=2000$ and $\rho \in \{0, 0.5, 1\}$. 
 }\label{CH2tab:NLE}
	\centering
% \resizebox{0.85\textwidth}{!}{
	\begin{tabular}{c||ccc|ccc|ccc}
		 & \multicolumn{3}{c|}{Bias} & \multicolumn{3}{c|}{MSE} & \multicolumn{3}{c}{Coverage} \\ \hline
		 $\rho$      & ATE    & SD     & PEP    & ATE    & SD     & PEP   & ATE      & SD      & PEP     \\ \hline
0.00 & 0.02 & 0.02 & 0.02 & 0.02 & 0.02 & 0.00 & 0.96 & 0.91 & 0.88 \\ 
  0.50 & 0.02 & 0.04 & 0.01 & 0.02 & 0.02 & 0.00 & 0.95 & 0.91 & 0.92 \\ 
  1.00 & 0.02 & 0.08 & -0.01 & 0.01 & 0.03 & 0.00 & 0.94 & 0.90 & 0.92 \\ 
	\end{tabular}%
% }
\end{table} 

\begin{figure}[H]
	\centering
	\resizebox{0.9\linewidth}{!}{\includegraphics{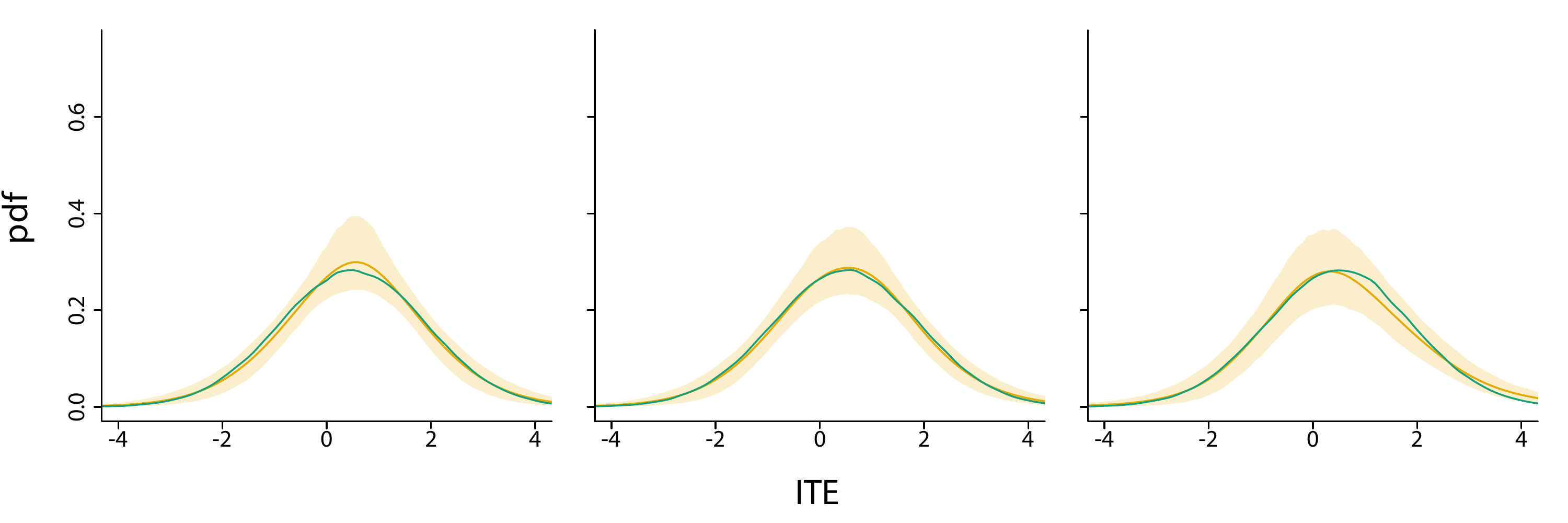}}
	\captionsetup{width=0.9\textwidth}
	\caption{Pointwise empirical mean (opaque yellow) as well as the $2.5\%$ and $97.5\%$ quantiles (transparent yellow) of the estimated ITE densities using the extended CRF of $1000$ simulations. The association of the measured $X_{0}$ and unmeasured modifier $U_{1}$ varies per column equal to $0, 0.50$ and $1$ (left to right), and the sample size per simulation equals $2000$. Furthermore, the actual ITE distribution is presented (green). }\label{CH2NLfig3}
\end{figure}

\subsection{Confounders that are no modifiers}\label{app:B4}
To emphasize that features might be confounders but no modifiers at the additive scale, we consider the cause-effect relations in Equation \eqref{eq:example}  using $\tau_{\text{sex}}=0$ so that sex is only a confounder. All other parameter values are taken the same as in the main example. 

 \begin{table}[H]
	\captionsetup{width=0.9\textwidth}
	\caption{Bias, mean squared error (MSE) and coverage of the estimated mean, SD and PEP ($\mathbb{P}(Y^1-Y^0>0)$) of the ITE distribution using the characteristics of the CATE distribution as an estimator based on $1000$ simulated samples for $n=2000$, and $\rho \in \{0, 0.5, 1\}$. 
 }\label{CH2tab:C1}
	\centering
% \resizebox{0.85\textwidth}{!}{
	\begin{tabular}{c||ccc|ccc|ccc}
		 & \multicolumn{3}{c|}{Bias} & \multicolumn{3}{c|}{MSE} & \multicolumn{3}{c}{Coverage} \\ \hline
   	 $\rho$      & ATE    & SD     & PEP    & ATE    & SD     & PEP   & ATE      & SD      & PEP     \\ \hline
0.00 & 0.01 & -1.18 & 0.34 & 0.02 & 1.38 & 0.12 & 0.94 & 0.00 & 0.19 \\ 
  0.50 & -0.00 & -0.83 & 0.15 & 0.02 & 0.69 & 0.03 & 0.96 & 0.00 & 0.59 \\ 
  1.00 & 0.00 & -0.20 & 0.01 & 0.01 & 0.05 & 0.00 & 0.95 & 0.44 & 0.94 \\ 
	\end{tabular}%
% }
\end{table}

\begin{figure}[H]
	\centering
 	\captionsetup{width=0.9\textwidth}
	\resizebox{0.9\linewidth}{!}{\includegraphics{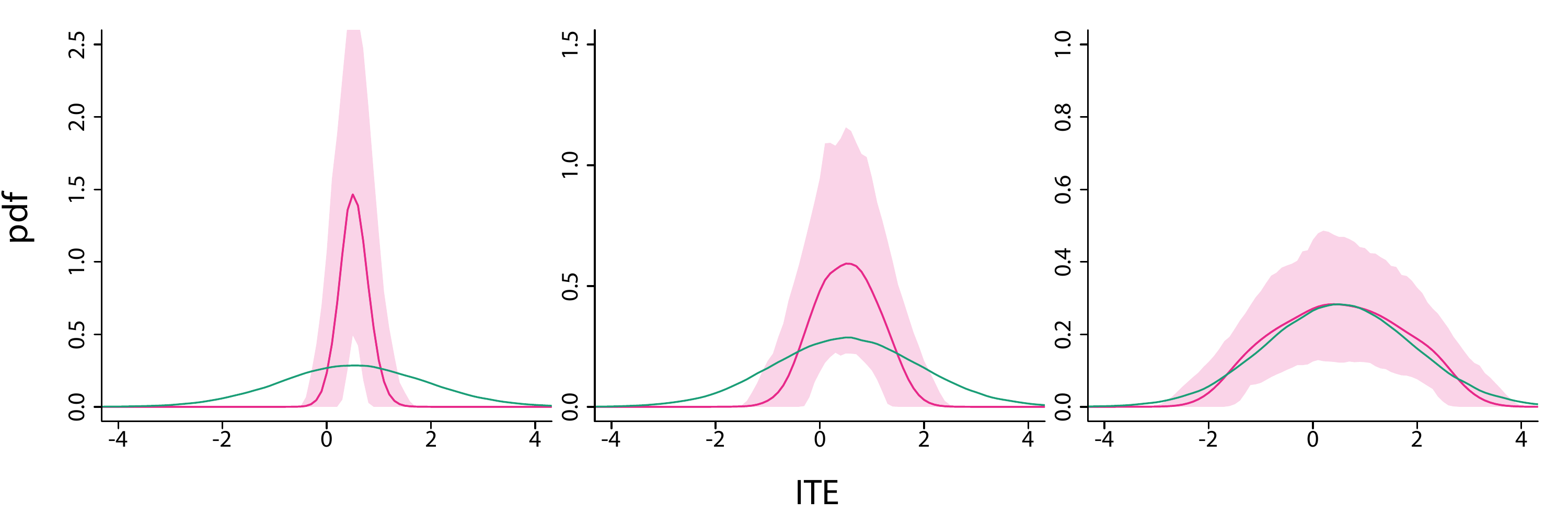}}
	\caption{Pointwise empirical mean (opaque pink) as well as the $2.5\%$ and $97.5\%$ quantiles (transparent pink) of the estimated ITE densities of $1000$ simulations. The association of the measured $X_{0}$ and unmeasured modifier $U_{1}$ varies per column equal to $0, 0.50$ and $1$ (left to right), and the sample size per simulation equals $2000$. Furthermore, the actual ITE distribution is presented (green). }\label{CH2C1fig2}
\end{figure}

\begin{table}[H]
	\captionsetup{width=0.9\textwidth}
	\caption{Bias, mean squared error (MSE) and coverage of the estimated mean, SD and PEP ($\mathbb{P}(Y^1-Y^0>0)$) of the ITE distribution using the extended CRF based on $1000$ simulated samples for $n=2000$ and $\rho \in \{0, 0.5, 1\}$. 
 }\label{CH2tab:C1E}
	\centering
% \resizebox{0.85\textwidth}{!}{
	\begin{tabular}{c||ccc|ccc|ccc}
		 & \multicolumn{3}{c|}{Bias} & \multicolumn{3}{c|}{MSE} & \multicolumn{3}{c}{Coverage} \\ \hline
		 $\rho$      & ATE    & SD     & PEP    & ATE    & SD     & PEP   & ATE      & SD      & PEP     \\ \hline
0.00 & 0.01 & 0.04 & 0.01 & 0.02 & 0.02 & 0.00 & 0.94 & 0.92 & 0.89 \\ 
  0.50 & -0.00 & 0.06 & 0.00 & 0.02 & 0.02 & 0.00 & 0.96 & 0.90 & 0.92 \\ 
  1.00 & 0.00 & 0.09 & -0.02 & 0.01 & 0.03 & 0.00 & 0.95 & 0.90 & 0.90 \\ 
	\end{tabular}%
% }
\end{table} 

\begin{figure}[H]
	\centering
	\resizebox{0.9\linewidth}{!}{\includegraphics{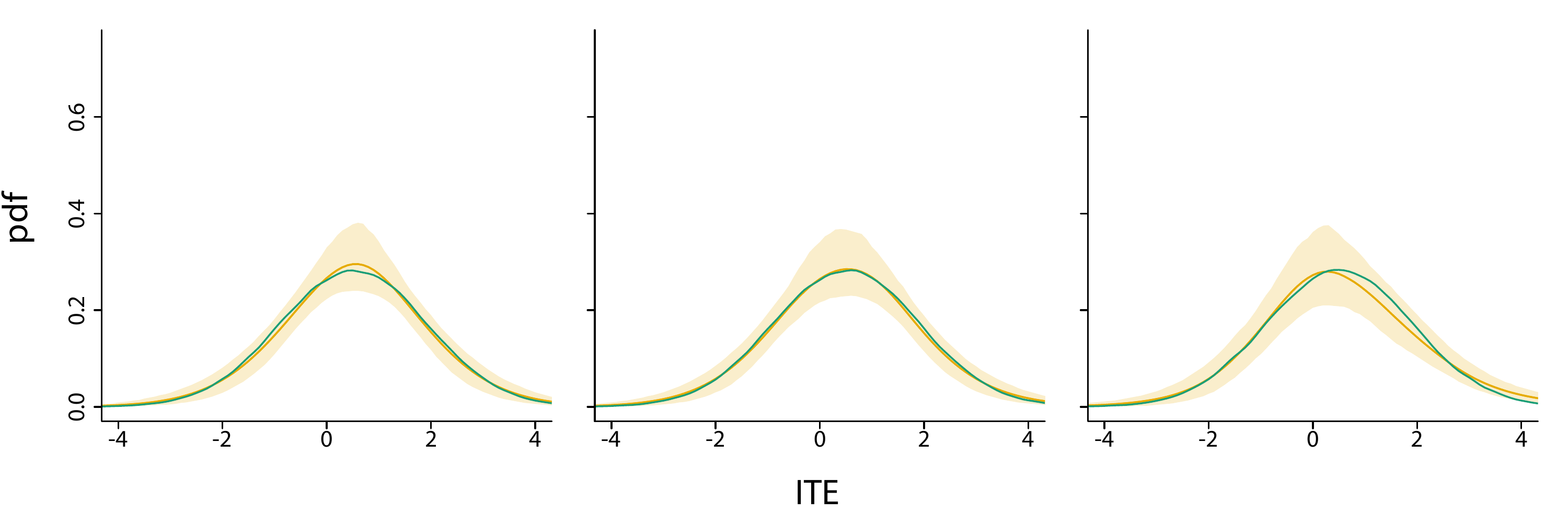}}
	\captionsetup{width=0.9\textwidth}
	\caption{Pointwise empirical mean (opaque yellow) as well as the $2.5\%$ and $97.5\%$ quantiles (transparent yellow) of the estimated ITE densities using the extended CRF of $1000$ simulations. The association of the measured $X_{0}$ and unmeasured modifier $U_{1}$ varies per column equal to $0, 0.50$ and $1$ (left to right), and the sample size per simulation equals $2000$. Furthermore, the actual ITE distribution is presented (green). }\label{CH2C1fig3}
\end{figure}

If $\mathbb{E}[Y~{\mid}~A{{=}}1, \boldsymbol{X}{=}\boldsymbol{x}] - \mathbb{E}[Y~{\mid}~A{{=}}0, \boldsymbol{X}{=}\boldsymbol{x}]$ is a complex function in $\boldsymbol{x}$, ML methods can be very usefull. However, to estimate CATEs appropriately, $\mathbb{E}[Y~{\mid}~A{{=}}a, \boldsymbol{X}{=}\boldsymbol{x}]$ should equal $\mathbb{E}[Y^a  \mid \boldsymbol{X}{=}\boldsymbol{x}]$. Thus, we should accurately account for confounding. When using the CRF to estimate CATEs, the R-learner decomposition facilitates this, as described in Section \ref{CH2sec:2.1}. In case this decomposition is not used, the CATE is estimated as
\begin{equation} \hat{{}\tau}(\boldsymbol{x}) ~{=}~ \frac{\sum_{i{=}1}^{n} \alpha_{i}(\boldsymbol{x}) Y_{i} A_{i} }{\sum_{i{=}1}^{n} \alpha_{i}(\boldsymbol{x}) A_{i} } - \frac{\sum_{i{=}1}^{n} \alpha_{i}(\boldsymbol{x})  Y_{i}  \left(1-A_{i} \right)}{\sum_{i{=}1}^{n} \alpha_{i}(\boldsymbol{x}) \left(1-A_{i} \right)}, \end{equation} where now also the similarity weights $\alpha_{i}(\boldsymbol{x})$ are obtained from a random forest fitted on the non-centralized outcome and treatment. So, confounders that are no effect modifiers (on the additive scale) will not be accounted for. This estimator is implemented in the \texttt{grf} package by using \texttt{causal\_forest(X,Y,W,Y.hat{=}0, W.hat{=}0)}. 

\begin{figure}[H]
		\captionsetup{width=0.9\textwidth}
	\centering
	\begin{subfigure}{.4\textwidth}
		%\centering
		\resizebox{1\textwidth}{!}{\includegraphics{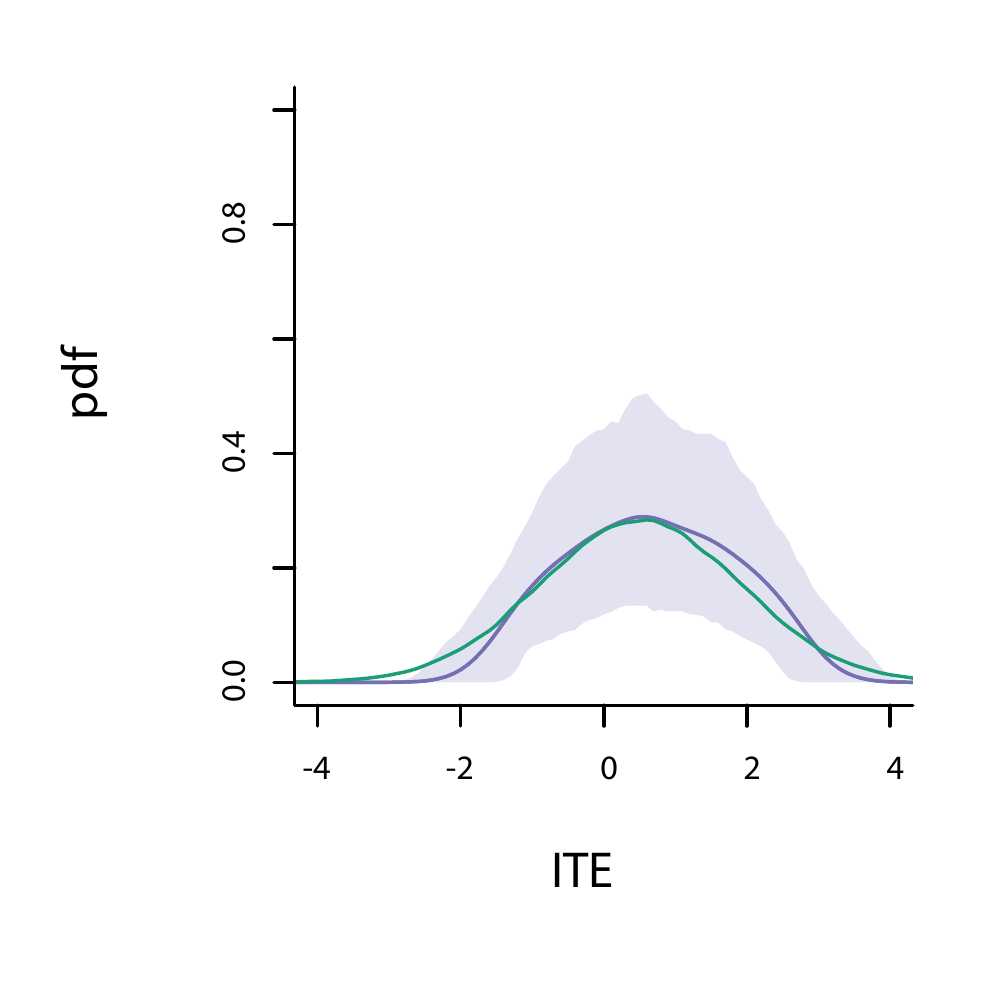}}\vspace{-0.5cm}
		\caption{}\label{CH2C1NOa}	
	\end{subfigure}
	\begin{subfigure}{.4\textwidth}
		%\centering
		\resizebox{1\textwidth}{!}{\includegraphics{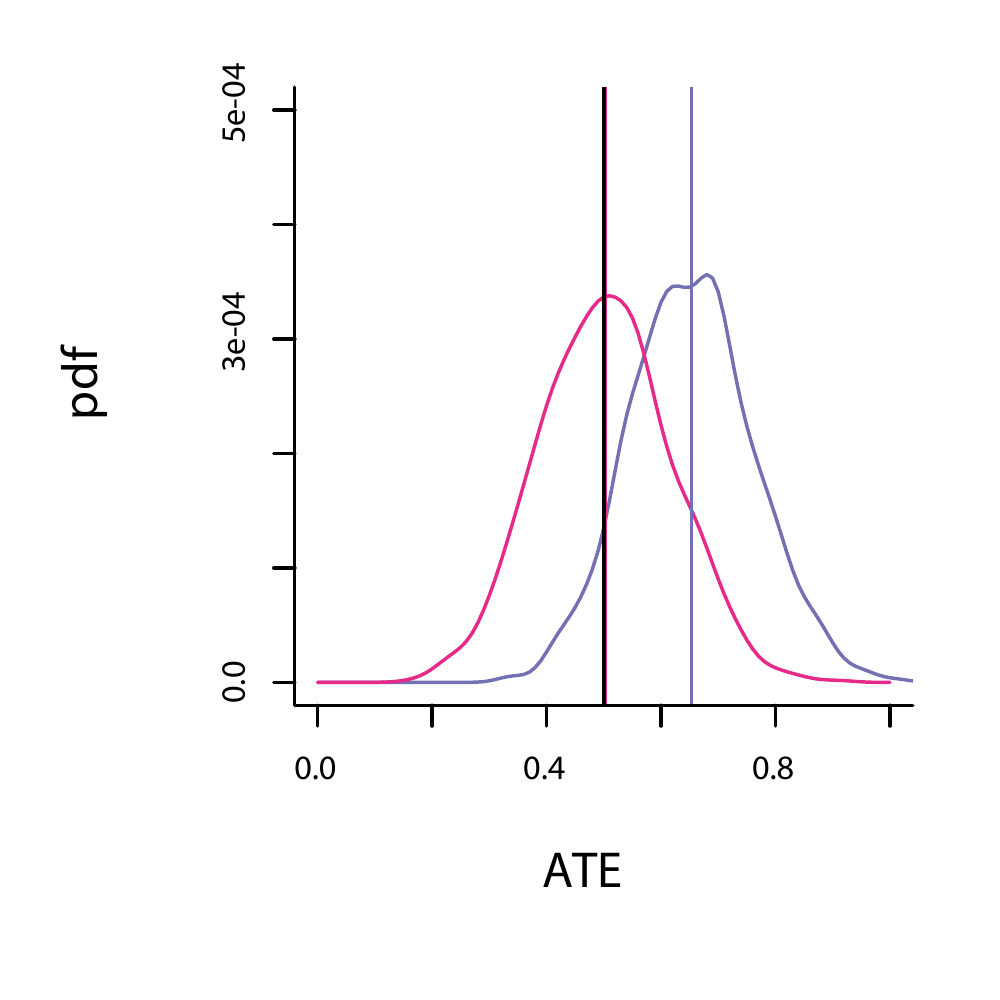}}\vspace{-0.5cm}
		\caption{}\label{CH2C1NOb}	
	\end{subfigure}
	\caption{ (a). Pointwise empirical mean (opaque blue), as well as the $2.5\%$ and $97.5\%$ quantiles (transparent blue) of the estimated ITE densities of $1000$ simulations using the CRF without orthogonalization together with the actual ITE distribution (green). (b). The distribution of the estimated ATE per simulation with (pink) and without the orthogonalization (blue). The average ATE over the $1000$ simulations and the actual ATE are presented with vertical lines.}\label{CH2C1NO}
\end{figure}

For the current setting, the ATE estimator based on the CRF without orthogonalization is biased, as illustrated in Table \ref{CH2tab:C1NO}. Since the bias is relatively small, for $\rho=1$, the ITE distribution based on the CRF is still close to the actual distribution as is presented in Figure \ref{CH2C1NOa}. However, the bias in the ATE is clearly visible when we consider the ATEs estimated for all simulations, as is shown in Figure \ref{CH2C1NOa}. The R-learner decomposition is necessary to account for the confounding by sex. 

 \begin{table}[H]
	\captionsetup{width=0.9\textwidth}
	\caption{Bias, mean squared error (MSE) and coverage of the estimated mean, SD and PEP ($\mathbb{P}(Y^1-Y^0>0)$) of the ITE distribution using the characteristics of the CATE distribution without using the orthogonalization as an estimator based on $1000$ simulated samples for $n=2000$, and $\rho \in \{0, 0.5, 1\}$. 
 }\label{CH2tab:C1NO}
	\centering
% \resizebox{0.85\textwidth}{!}{
	\begin{tabular}{c||ccc|ccc|ccc}
		 & \multicolumn{3}{c|}{Bias} & \multicolumn{3}{c|}{MSE} & \multicolumn{3}{c}{Coverage} \\ \hline
   	 $\rho$      & ATE    & SD     & PEP    & ATE    & SD     & PEP   & ATE      & SD      & PEP     \\ \hline
1.00 & 0.15 & -0.22 & 0.04 & 0.04 & 0.06 & 0.00 & 0.15 & 0.55 & 0.88 \\
\end{tabular}%
% }
\end{table}

\begin{figure}[H]
		\captionsetup{width=0.9\textwidth}
	\centering
	\begin{subfigure}{.4\textwidth}
		%\centering
		\resizebox{1\textwidth}{!}{\includegraphics{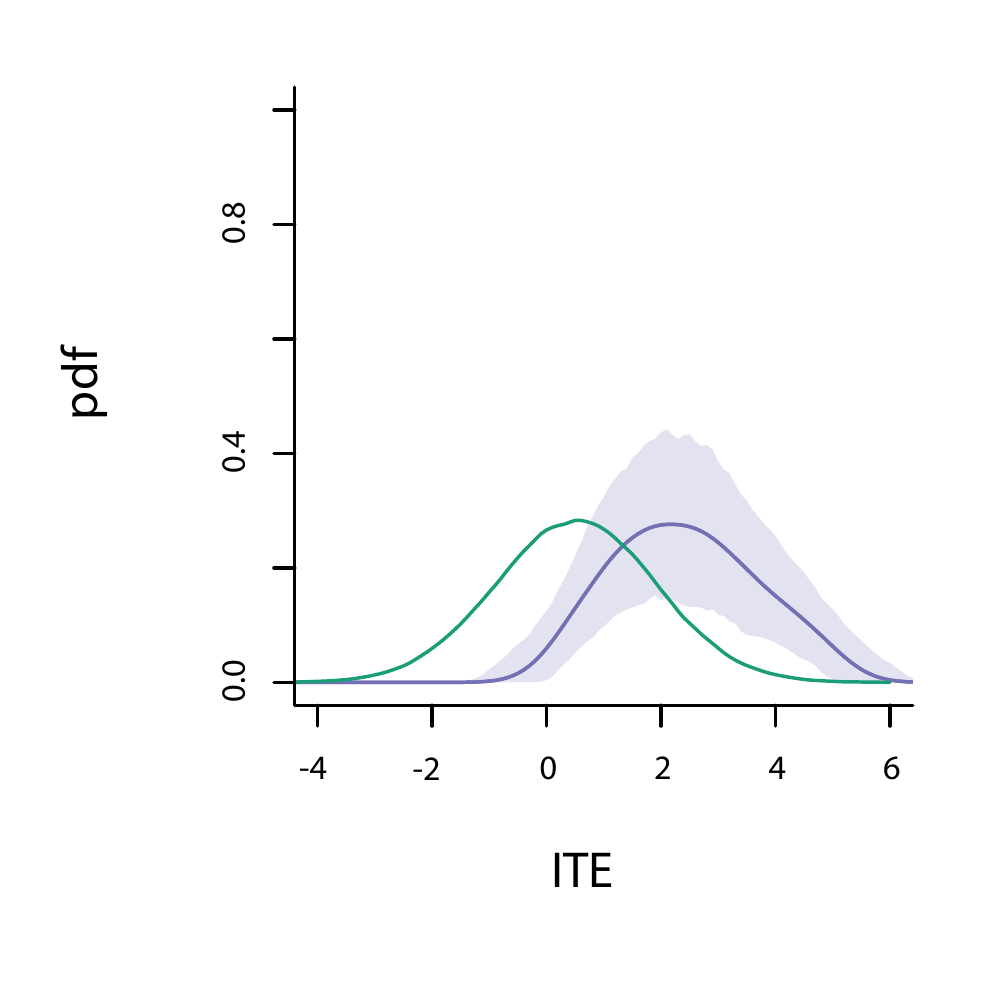}}\vspace{-0.5cm}
		\caption{}\label{CH2C2NOa}	
	\end{subfigure}
	\begin{subfigure}{.4\textwidth}
		%\centering
		\resizebox{1\textwidth}{!}{\includegraphics{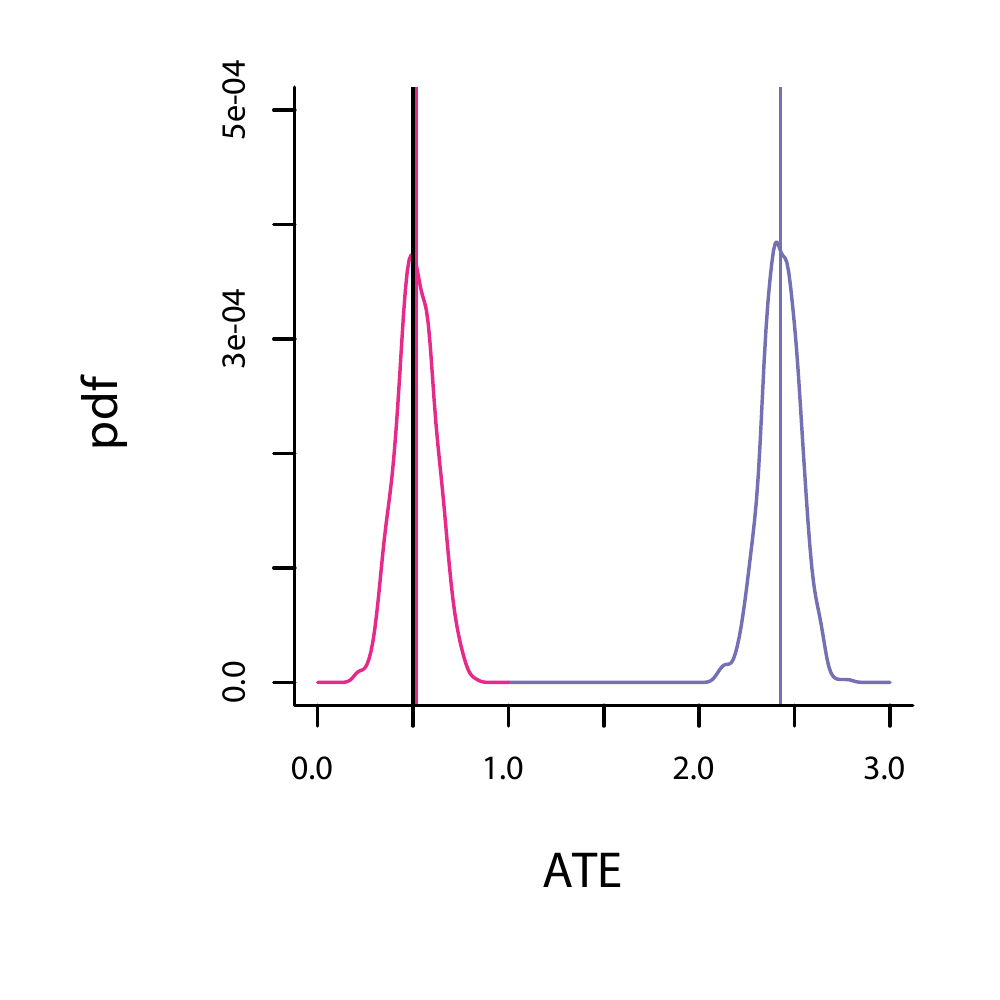}}\vspace{-0.5cm}
		\caption{}\label{CH2C2NOb}	
	\end{subfigure}
	\caption{ (a). Pointwise empirical mean (opaque blue), as well as the $2.5\%$ and $97.5\%$ quantiles (transparent blue) of the estimated ITE densities of $1000$ simulations using the CRF without orthogonalization together with the actual ITE distribution (green). (b). The distribution of the estimated ATE per simulation with (pink) and without the orthogonalization (blue). The average ATE over the $1000$ simulations and the actual ATE are presented with vertical lines.}\label{CH2C2NO}
\end{figure}

\noindent By additionally changing $\beta_{\text{sex}}=3.2$ and $\alpha_{\text{sex}}=3$, sex becomes a much stronger confounder, and the problem becomes even more apparent, as shown in Table \ref{CH2tab:C2NO} and Figure \ref{CH2C2NO}.

 \begin{table}[H]
	\captionsetup{width=0.9\textwidth}
	\caption{Bias, mean squared error (MSE) and coverage of the estimated mean, SD and PEP ($\mathbb{P}(Y^1-Y^0>0)$) of the ITE distribution using the characteristics of the CATE distribution without using the orthogonalization as an estimator based on $1000$ simulated samples for $n=2000$, and $\rho \in \{0, 0.5, 1\}$. 
 }\label{CH2tab:C2NO}
	\centering
% \resizebox{0.85\textwidth}{!}{
	\begin{tabular}{c||ccc|ccc|ccc}
		 & \multicolumn{3}{c|}{Bias} & \multicolumn{3}{c|}{MSE} & \multicolumn{3}{c}{Coverage} \\ \hline
   	 $\rho$      & ATE    & SD     & PEP    & ATE    & SD     & PEP   & ATE      & SD      & PEP     \\ \hline
1.00 & 1.93 & -0.11 & 0.34 & 3.72 & 0.02 & 0.12 & 0.00 & 0.90 & 0.00 \\
\end{tabular}%
% }
\end{table}

When interested in the heterogeneity of effects, the objective function of an ML method is typically focused on maximizing the heterogeneity in differences between observed treated and controls as a function of the features ($\boldsymbol{X}$) at hand. As a result, the method might not account for confounders that are no effect modifiers (on the additive scale), and biased causal effects may be obtained.

\subsection{Absence of confounding}\label{app:B5}
Finally, we considered a scenario with the absence of confounding by adjusting the cause-effect relations in Equation \eqref{eq:example} by
$$A\sim \text{ber}(0.15),$$
so that the fraction of exposed individuals is similar to the main example. In this scenario, sex and SBP are only modifiers but no confounders. All other parameters were taken the same as in the main example.

 \begin{table}[H]
	\captionsetup{width=0.9\textwidth}
	\caption{Bias, mean squared error (MSE) and coverage of the estimated mean, SD and PEP ($\mathbb{P}(Y^1-Y^0>0)$) of the ITE distribution using the characteristics of the CATE distribution as an estimator based on $1000$ simulated samples for $n=2000$, and $\rho \in \{0, 0.5, 1\}$. 
 }\label{CH2tab:PM}
	\centering
% \resizebox{0.85\textwidth}{!}{
	\begin{tabular}{c||ccc|ccc|ccc}
		 & \multicolumn{3}{c|}{Bias} & \multicolumn{3}{c|}{MSE} & \multicolumn{3}{c}{Coverage} \\ \hline
   	 $\rho$      & ATE    & SD     & PEP    & ATE    & SD     & PEP   & ATE      & SD      & PEP     \\ \hline
0.00 & 0.01 & -1.18 & 0.33 & 0.02 & 1.39 & 0.11 & 0.96 & 0.00 & 0.17 \\ 
  0.50 & 0.01 & -0.82 & 0.14 & 0.02 & 0.69 & 0.03 & 0.95 & 0.00 & 0.61 \\ 
  1.00 & -0.00 & -0.22 & 0.00 & 0.01 & 0.06 & 0.00 & 0.96 & 0.40 & 0.95 \\  
	\end{tabular}%
% }
\end{table}

\begin{figure}[H]
	\centering
 	\captionsetup{width=0.9\textwidth}
	\resizebox{0.9\linewidth}{!}{\includegraphics{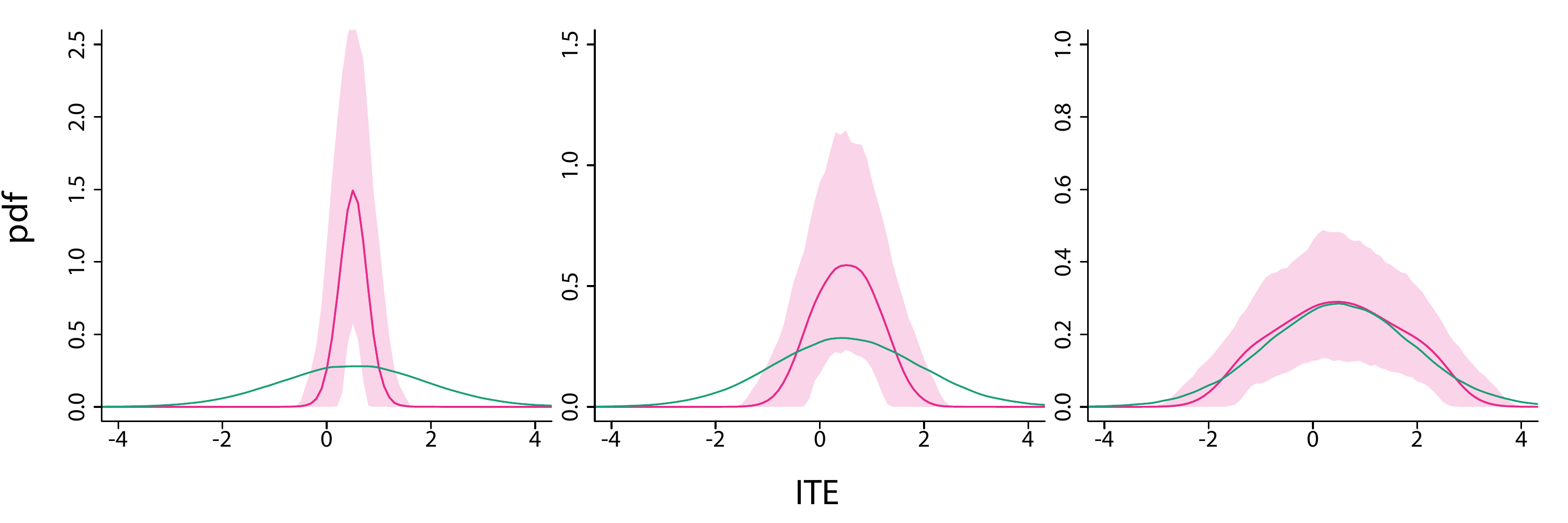}}
	\caption{Pointwise empirical mean (opaque pink) as well as the $2.5\%$ and $97.5\%$ quantiles (transparent pink) of the estimated ITE densities of $1000$ simulations. The association of the measured $X_{0}$ and unmeasured modifier $U_{1}$ varies per column equal to $0, 0.50$ and $1$ (left to right), and the sample size per simulation equals $2000$. Furthermore, the actual ITE distribution is presented (green). }\label{CH2PMfig2}
\end{figure}

\begin{table}[H]
	\captionsetup{width=0.9\textwidth}
	\caption{Bias, mean squared error (MSE) and coverage of the estimated mean, SD and PEP ($\mathbb{P}(Y^1-Y^0>0)$) of the ITE distribution using the extended CRF based on $1000$ simulated samples for $n=2000$ and $\rho \in \{0, 0.5, 1\}$. 
 }\label{CH2tab:PME}
	\centering
% \resizebox{0.85\textwidth}{!}{
	\begin{tabular}{c||ccc|ccc|ccc}
		 & \multicolumn{3}{c|}{Bias} & \multicolumn{3}{c|}{MSE} & \multicolumn{3}{c}{Coverage} \\ \hline
		 $\rho$      & ATE    & SD     & PEP    & ATE    & SD     & PEP   & ATE      & SD      & PEP     \\ \hline
0.00 & 0.01 & 0.03 & 0.01 & 0.02 & 0.02 & 0.00 & 0.96 & 0.92 & 0.92 \\ 
  0.50 & 0.01 & 0.03 & 0.01 & 0.02 & 0.02 & 0.00 & 0.95 & 0.92 & 0.92 \\ 
  1.00 & -0.00 & 0.05 & -0.01 & 0.01 & 0.02 & 0.00 & 0.96 & 0.93 & 0.94 \\ 
	\end{tabular}%
% }
\end{table} 

\begin{figure}[H]
	\centering
	\resizebox{0.9\linewidth}{!}{\includegraphics{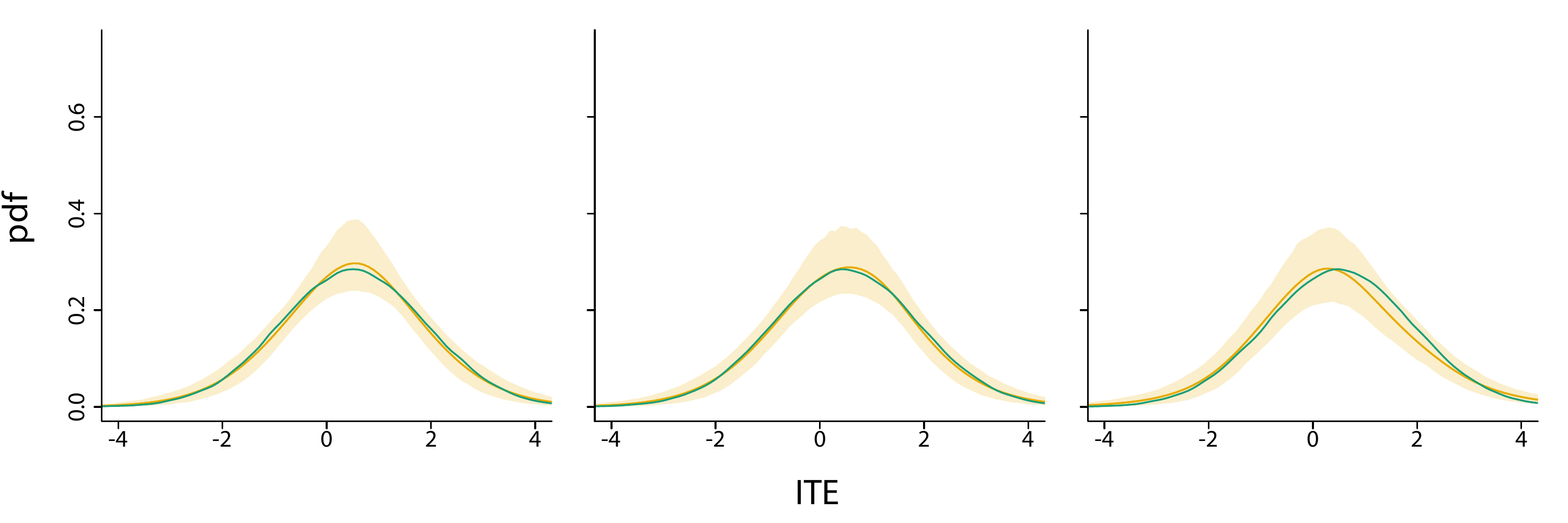}}
	\captionsetup{width=0.9\textwidth}
	\caption{Pointwise empirical mean (opaque yellow) as well as the $2.5\%$ and $97.5\%$ quantiles (transparent yellow) of the estimated ITE densities using the extended CRF of $1000$ simulations. The association of the measured $X_{0}$ and unmeasured modifier $U_{1}$ varies per column equal to $0, 0.50$ and $1$ (left to right), and the sample size per simulation equals $2000$. Furthermore, the actual ITE distribution is presented (green). }\label{CH2PMfig3}
\end{figure}

\begin{figure}[H]
		\captionsetup{width=0.9\textwidth}
	\centering
	\begin{subfigure}{.4\textwidth}
		%\centering
		\resizebox{1\textwidth}{!}{\includegraphics{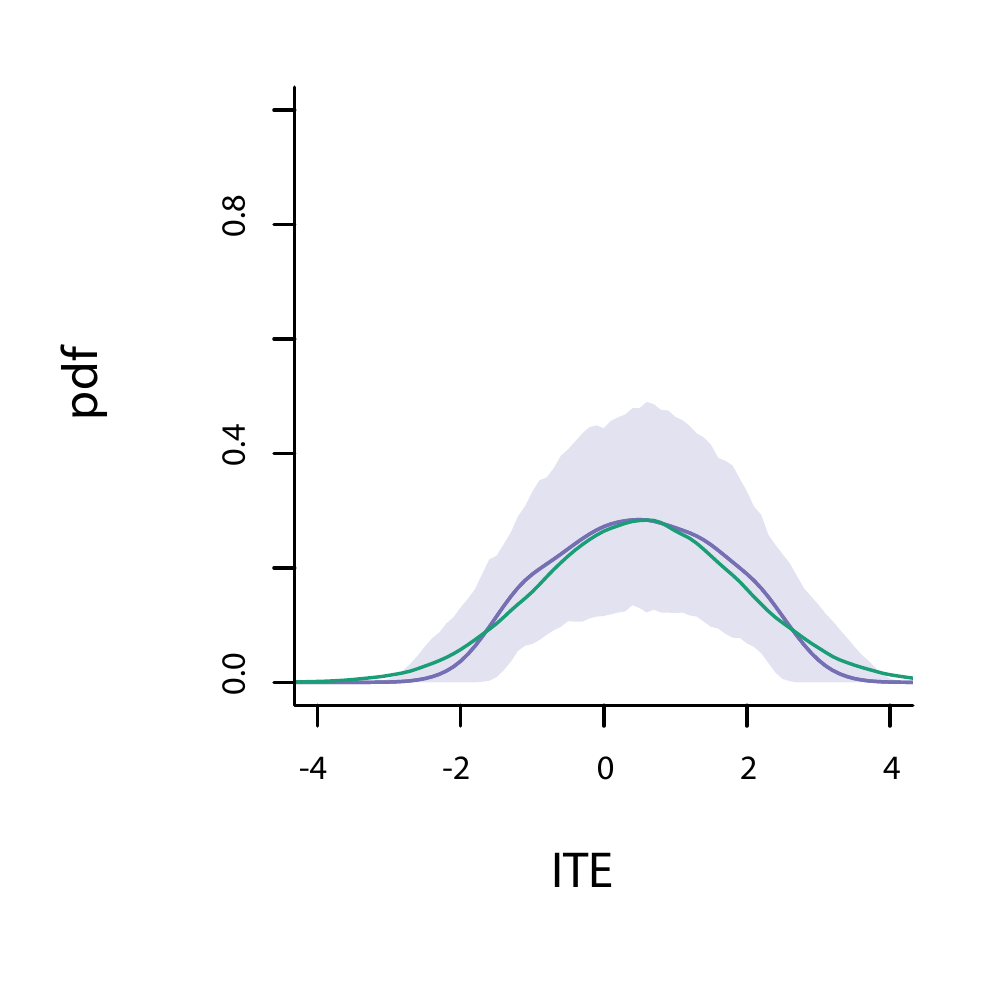}}\vspace{-0.5cm}
		\caption{}\label{CH2PMNOa}	
	\end{subfigure}
	\begin{subfigure}{.4\textwidth}
		%\centering
		\resizebox{1\textwidth}{!}{\includegraphics{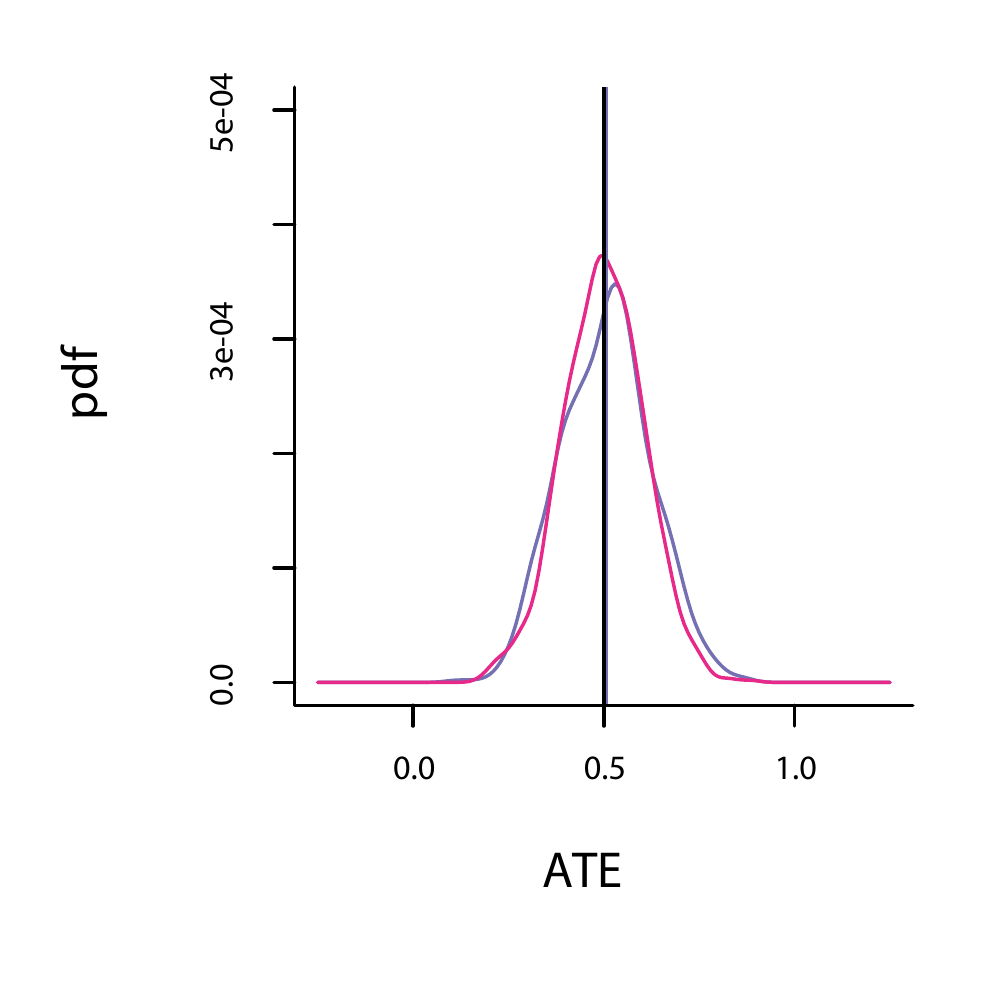}}\vspace{-0.5cm}
		\caption{}\label{CH2PMNOb}	
	\end{subfigure}
	\caption{ (a). Pointwise empirical mean (opaque blue), as well as the $2.5\%$ and $97.5\%$ quantiles (transparent blue) of the estimated ITE densities of $1000$ simulations using the CRF without orthogonalization together with the actual ITE distribution (green). (b). The distribution of the estimated ATE per simulation with (pink) and without the orthogonalization (blue). The average ATE over the $1000$ simulations and the actual ATE are presented with vertical lines.}\label{CH2PMNO}
\end{figure}

For completeness, contrary to the cases with confounding, in this scenario, the R-learner decomposition is not necessary to estimate the CATEs as illustrated in Table \ref{CH2tab:PMNO} and Figure \ref{CH2PMNO}. 

 \begin{table}[H]
	\captionsetup{width=0.9\textwidth}
	\caption{Bias, mean squared error (MSE) and coverage of the estimated mean, SD and PEP ($\mathbb{P}(Y^1-Y^0>0)$) of the ITE distribution using the characteristics of the CATE distribution without using the orthogonalization as an estimator based on $1000$ simulated samples for $n=2000$. 
 }\label{CH2tab:PMNO}
	\centering
% \resizebox{0.85\textwidth}{!}{
	\begin{tabular}{c||ccc|ccc|ccc}
		 & \multicolumn{3}{c|}{Bias} & \multicolumn{3}{c|}{MSE} & \multicolumn{3}{c}{Coverage} \\ \hline
   	 $\rho$      & ATE    & SD     & PEP    & ATE    & SD     & PEP   & ATE      & SD      & PEP     \\ \hline
1.00 & 0.01 & -0.22 & 0.00 & 0.01 & 0.06 & 0.00 & 0.35 & 0.58 & 0.96 \\ 
\end{tabular}%
% }
\end{table}

%TC:endignore
\end{document}